\documentclass[aps,epsf,floatfix,prl,twocolumn]{revtex4-1}

\usepackage{graphics,graphicx,epsfig}
\usepackage{amssymb,color}
\usepackage{epsf,epstopdf,wrapfig}
\usepackage{mciteplus}
\usepackage{graphicx,color}
\usepackage{amsmath,amssymb}
\usepackage{dcolumn}   % needed for some tables
\usepackage{bm}        % for https://fr.overleaf.com/project/62e245c5b6997cbe566767d3math
\usepackage{multirow}
\usepackage{dsfont}
\usepackage{lipsum,afterpage}
\usepackage{soul}
\usepackage{xcolor}
\usepackage{enumitem}

\def\be{\begin{equation}}
\def\ee{\end{equation}}
\def\bea{\begin{eqnarray}}
\def\eea{\end{eqnarray}}

\def\ra{\rangle}
\def\la{\langle}

\newcommand{\fref}[1]{Fig.~\ref{#1}} %
\newcommand{\aref}[1]{\hyperref[#1]{Appendix~\ref{#1}}}

\begin{document}

\title{Role of cilia activity and surrounding viscous fluid on properties of metachronal waves}

\author{Supravat Dey}
\email{supravat.dey@gmail.com}
\affiliation{Department of Physics and Department Computer Science and Engineering, SRM University - AP, Amaravati, Andhra Pradesh 522240}
\author{Gladys Massiera}
\email{gladys.massiera@umontpellier.fr}
\affiliation{ Laboratoire Charles Coulomb (L2C), Universit\'e de Montpellier, CNRS, 34095 Montpellier, France \\}
\author{Estelle Pitard}
\email{estelle.pitard@umontpellier.fr}
\affiliation{ Laboratoire Charles Coulomb (L2C), Universit\'e de Montpellier, CNRS, 34095 Montpellier, France \\}

\begin{abstract}
 Large groups of active cilia collectively beat in a fluid medium as metachronal waves, essential for some microorganisms motility and for flow generation in mucociliary clearance. Several models can predict the emergence of metachronal waves, but what controls the properties of metachronal waves is still unclear. Here, we investigate numerically  the respective impacts of active beating and viscous dissipation on the properties of metachronal waves in a collection of oscillators, using a simple  model for cilia in the presence of noise on regular lattices in one- and two-dimensions. We characterize the wave using spatial correlation and the frequency of collective beating. Our results clearly show that the viscosity of the fluid medium does not affect the wavelength; the activity of the cilia does. These numerical results are supported by a dimensional analysis, which shows that the result of wavelength invariance is robust against the model taken for sustained beating and the structure of hydtodynmanic coupling.  Interestingly, enhancement of cilia activity increases the wavelength and decreases the beating frequency, keeping the wave velocity almost unchanged. These results might have significance in understanding paramecium locomotion and mucociliary clearance diseases. 

\end{abstract}

\maketitle

\section{I. Introduction}
The emergence of phase-travelling waves in dense arrays of active beating cilia, known as metachronal waves, is a complex multiscale physics problem \cite{Gray1928,Sleigh1962,Blake1972,Gilpin2020,Milana2020,Brumley2015,Zhang2021} and is nonequilibrium because of the internal activity-driven movements of cilia. The active beating of each cilium  arises from the sliding of microtubules by thousands of molecular motors, and the subsequent interaction with the surrounding fluid medium. The coupling of a large number of these oscillators lead to synchronized dynamics over larger length scales.  Illustrations are abundant in nature with ciliary living systems differing by cilia assembly geometry, cilia activity, or properties of the surrounding fluid. In respiratory tissues, the continuous cleaning of our lungs is provided by cilia beating waves that generate mucus flow \cite{Wanner1996,SmithRes08}. For certain microorganisms such as paramecium, synchronized beating of cilia help in their efficient locomotion \cite{Lauga2020}. The complexity of cilia active beating pattern and their interaction with each other through a complex environment makes it difficult to predict the emergent wave properties, despite recent theoretical and experimental advancements.

Models of cilia arrays \cite{CosentinoPre03,Elgeti2013,GolestanianRev11,Uchida2017,Chakrabarti2022}, aim to identify the conditions required for such a coordinated state and to comprehend the physical parameters that govern the properties of the metachronal wave and the subsequent mucus transport. Several models have been proposed \cite{Niedermayer2008,CosentinoPre03,GolestanianRev11,Elgeti2013,Solovev2022}, wherein the coupling is primarily described as a viscous hydrodynamic coupling. In these models, different types of active forces - from simple to complex, successfully generate continuous beating of a cilium. Numerical simulations enable to investigate the intricate structure of cilia by considering their beating as a filament bending wave \cite{Elgeti2013,LeoniPRE2012,Guirao2007}.  Another approach is to model cilia by actuated micron-sized beads called rotors \cite{Uchida2011,Brumley2016,Meng2021,Kanale2022} or rowers 
\cite{CosentinoPre03,CicutaPnas10,Hamilton2021}. For a large group of cilia arranged in an array, it has been shown that hydrodynamic coupling can lead to metachronal waves for various models of cilia \cite{Elgeti2013,Uchida2017,Chakrabarti2022}. 
Recently, the influence, on these collective behaviors, of several physical parameters such as noise \cite{Solovev2022b,Hamilton2021} and disorder in the arrangement and orientation of cilia has been investigated both numerically \cite{Deypre2018,Hamilton2021,Ramirez2020} and experimentally \cite{Pellicciotta2020,Ramirez2020}, showing that spatial heterogeneity favors transport. Other important physical quantities that may play a role on the coordination are the activity related to the sustained beating and the dissipation in the viscous fluid, that will have opposite impacts on the metachronal waves emerging from cilia beating. Experimentally, a decrease in beating frequency with viscosity was found \cite{Machemer1972,Johnson1991}, whereas the beating amplitude and the metachronal wavelength were found constant up to $\approx 50$ times the viscosity of water \cite{Kikuchi2017,Gheber1998}. Theoretically, the mutual influence of activity and dissipation were almost not explored \cite{Gheber1990}. Here, our fundamental inquiry pertains to the interplay between cilia beating characteristics and fluid medium and its impact on the overall properties of metachronal waves.

\section{II. The Rower model}
\begin{center}
\begin{figure}[t]
 \includegraphics[width=0.45\textwidth]{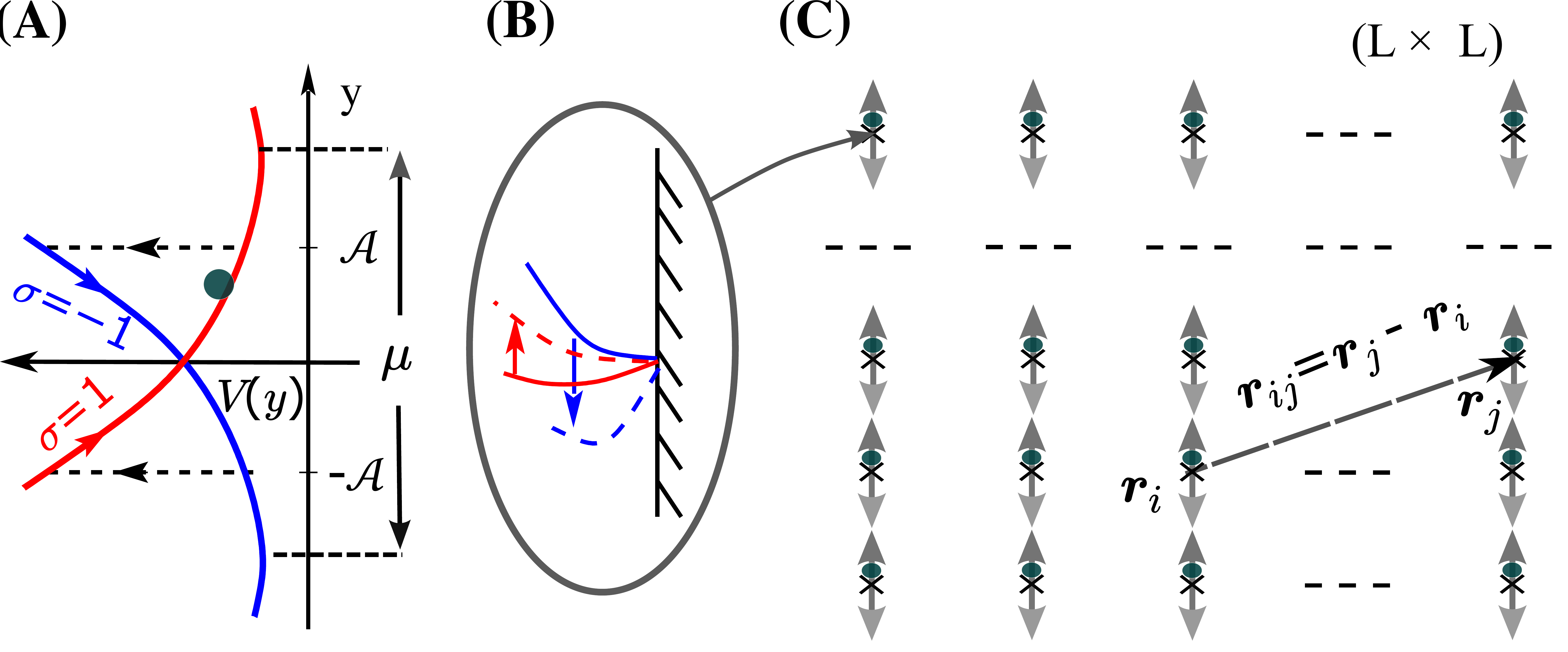}
\caption{Rower model of cilia \cite{CosentinoPre03}. (A) The motion of a micron-sized bead in a viscous medium under two harmonic potential branches, corresponding to $\sigma \pm 1$, represents the stroke and anti-stroke beating of a cilium. (B) The bead switches branches when it reaches terminal position $y=\pm \mathcal{A}$.  
(C) \textcolor{red}{Rowers are arranged in 2D on a $L\times L$ square lattice, with their beating direction along the $y$-axis, as indicated by the double-arrow.} Hydrodynamic interaction between rower $i$ and $j$ is modelled by the Oseen coupling, which is dependent on the $r_{ij}$ vector.}
\label{fig:themodel}
\end{figure}
\end{center}

We study the metachronal waves in the rower model of cilia in viscous fluid for one and two-dimensional regular lattices in the presence of thermal noise. In the rower model \cite{CosentinoPre03, CicutaPnas10, Hamilton2021}, the complex active beating of a cilium is simplified into the back and forth motion, along an axis, of a micron-sized bead immersed in a viscous fluid, thus ensuring a low Reynolds number regime.
Such an oscillating motion is driven by two harmonic potential branches, corresponding to the stroke and anti-stroke of the cilia beating, with a geometric switching mechanism. The bead moves downhill of a potential until it reaches one of the two terminal positions for which switching to the second branch occurs (\fref{fig:themodel}A, B). This switching is like pumping energy to lift the bead on the upper side of the other potential at the terminal. At a given time, the bead can be found in one of these two states, the stroke and anti-stroke of the cilia beating, represented by a discrete $\sigma=\pm1$. The driving force for a bead displacement $y$ for a given $\sigma$ can be written as
\bea
f(y,\sigma) &=-\frac{d V(y,\sigma)}{dy}&= -k(y-\sigma \mu/2),
\eea
where $k$ is the force constant associated with the harmonic potentials, $\mu$ is the distance between minima of two potentials, and $\mathcal{A}$ is the beating amplitude. The supply of energies during each downhill motion in a harmonic potential, $k\mathcal{A}^2/2$, and during each switch,  the pumping energy $k \mu \mathcal{A}$, keep the bead oscillating in the dissipating media. \textcolor{red}{We refer to this pumping energy during each switch as `activity' for the rower model of cilia.} Therefore, for a given $\mu$, the `activity' of the bead depends on values of $k$ and $\mathcal{A}$. Because of its simplicity and ability to capture the two-stroke beating of cilia, the rower model has become a method of choice for theoretical and experimental studies of synchronization in ciliary systems \cite{Chakrabarti2022,Pellicciotta2020}.

We consider a system of $N$ rowers beating in the $y$ direction in a viscous medium. Rowers are placed regularly in one- or two-dimensions (square) lattices  (see \fref{fig:themodel}C) at fixed positions $\bf{r}_i$ (for $i=\{1,2,3,...,N\})$. The displacement, $y_i$, of a rower $i$ is hydrodynamically coupled with the others and is given by
\bea
\frac{dy_i}{dt} = \frac{f_i}{\gamma} + \sum_{j\neq i} O(i,j)\,  f_j +  \xi_i, 
\label{eqn:dyn_coupled}
\eea
where $\gamma=6\pi\eta a$ is the viscous drag coefficient for a bead with radius $a$ and $O(i,j)$ the coupling strength between rower $i$ and $j$.  In the far-field hydrodynamic coupling approximation, for which both the distance from the surface and the distance between two adjacent rowers (lattice spacing $\ell$) are  large compared to $a$, $O(i,j)$ is set by the Oseen tensor: $O(i,j) = \frac{1}{8\pi\eta r_{ij}}  \left(1+(\frac{y_{ij}}{{r}_{ij}})^2 \right)$, with $i\neq j$, and ${\bf r}_{ij}={\bf r}_j -{\bf r}_i$, the separation vector between rowers $i$ and $j$. The last term is due to the thermal noise, obeying the following delta-correlation: $\langle \xi (t) \rangle =0, \text {  } \langle \xi (t_1) \xi (t_2) \rangle = 2 \,D \, \delta(t_1 -t_2)$. For simplicity, we assume no correlation between the noise acting on each of the rowers as in \cite{Hamilton2021}. The noise strength or diffusivity is equal to $D=k_B\,T/\gamma$, $k_B$ and $T$ being the Boltzmann constant and the temperature.
The displacement of a single isolated bead shows sustained oscillations with the frequency $\nu_0 = 1/(2\,\tau_d \log\left[(\mu + 2 \mathcal{A})/(\mu - 2 \mathcal{A})  \right])$, where $\tau_d= \gamma/k$ is the relaxation time for the bead to reach equilibrium in a harmonic potential \cite{CicutaPnas10}. Such two coupled rowers beat collectively with antiphase synchronization \cite{CosentinoPre03}.
For many rowers, the interplay between the activity of the rowers and coupling through the medium generates metachronal waves \cite{CosentinoPre03}.
\begin{figure}[t]
\includegraphics[width=0.45\textwidth]{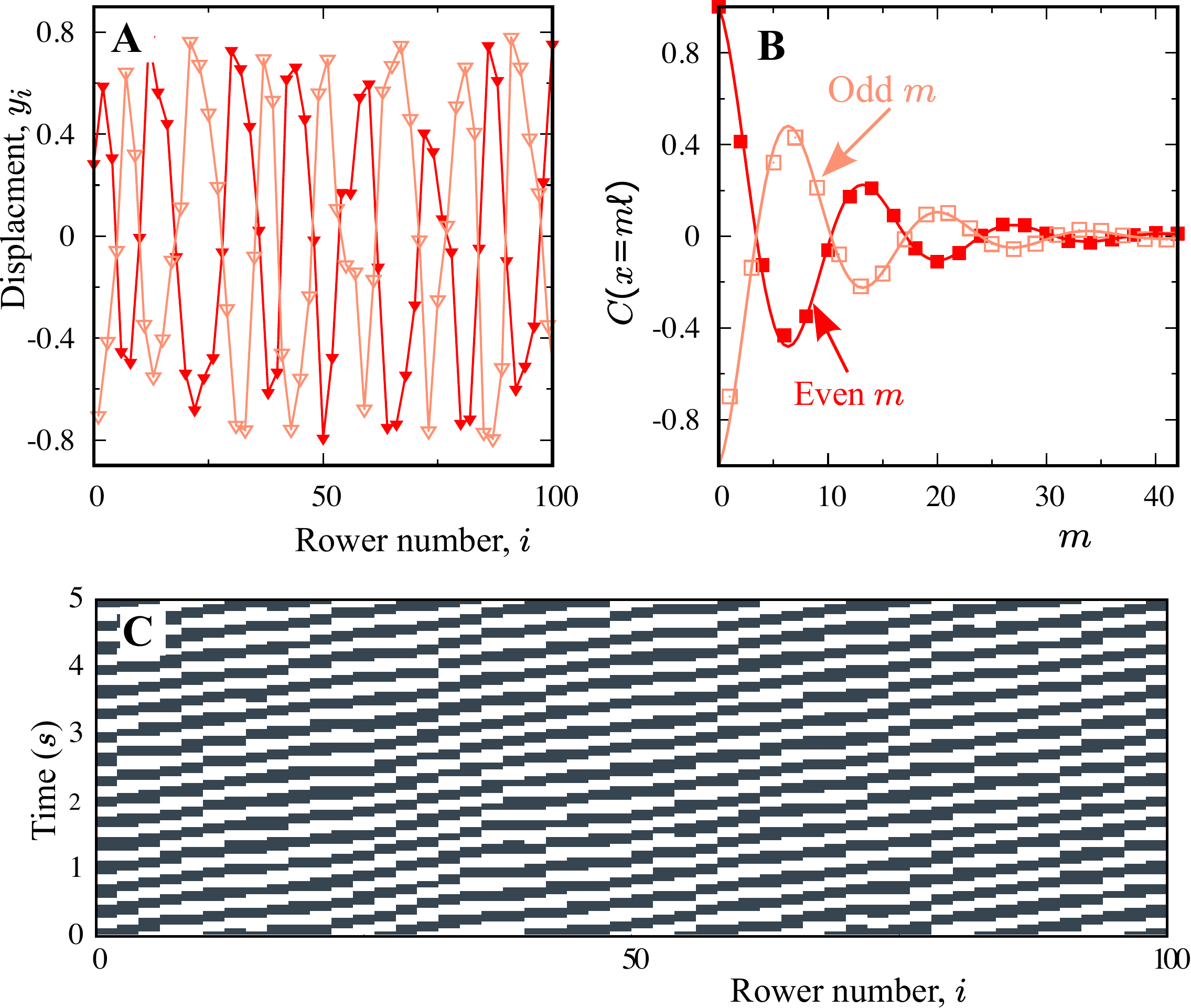}
\caption{Metachronal waves in 1d. A. Snapshot of displacements of the first 100 rowers. The displacement of even (odd) sites is plotted in light (dark) color. B. Correlation function $C(x{=}m\ell)$ between two rowers is plotted against separation distance $x$. C. Kymograph of the beating state $\sigma$,  with white color (black) representing $\sigma{=}1$ ($\sigma{=-1}$). Parameters: $\mathcal{A}=0.8$, $\eta{=}6\, \rm{mPa{\cdot}s}$, and $N{=}200$.}
\label{fig:fig2}
\end{figure}

\section{III. Numerical Results}

The Euler method with an integration step equal to $5{\times}10^{-3} s$  has been used to evolve the coupled dynamical equation (Eq.~\ref{eqn:dyn_coupled}), starting from random initial values for $\{\sigma_i, y_i\}$. The open boundary condition is implemented. Parameters are chosen within the experimentally relevant range \cite{CicutaPnas10,Brumley2012}: $a{=}1.5 \,\mu m$, $\ell{=}8\,\mu m$, 
 $k{=}2.6\,pN{\cdot}\mu m^{-1}$, $\mu{=}2\,\mu m$, $\mathcal{A}{=}0.56-0.8\,\mu m $, $\eta{=}2-20\, mPa{\cdot} s$, and $T{=}300K$. Results presented here for large system sizes; $N{=}L{=}200$ (for 1d) and $N{=}L^2{=}1600$ (for 2d). Comparing results with smaller systems (not shown here), we confirm that the presented results have no system size dependence.  
\begin{figure}[t]
\includegraphics[width=0.45\textwidth]{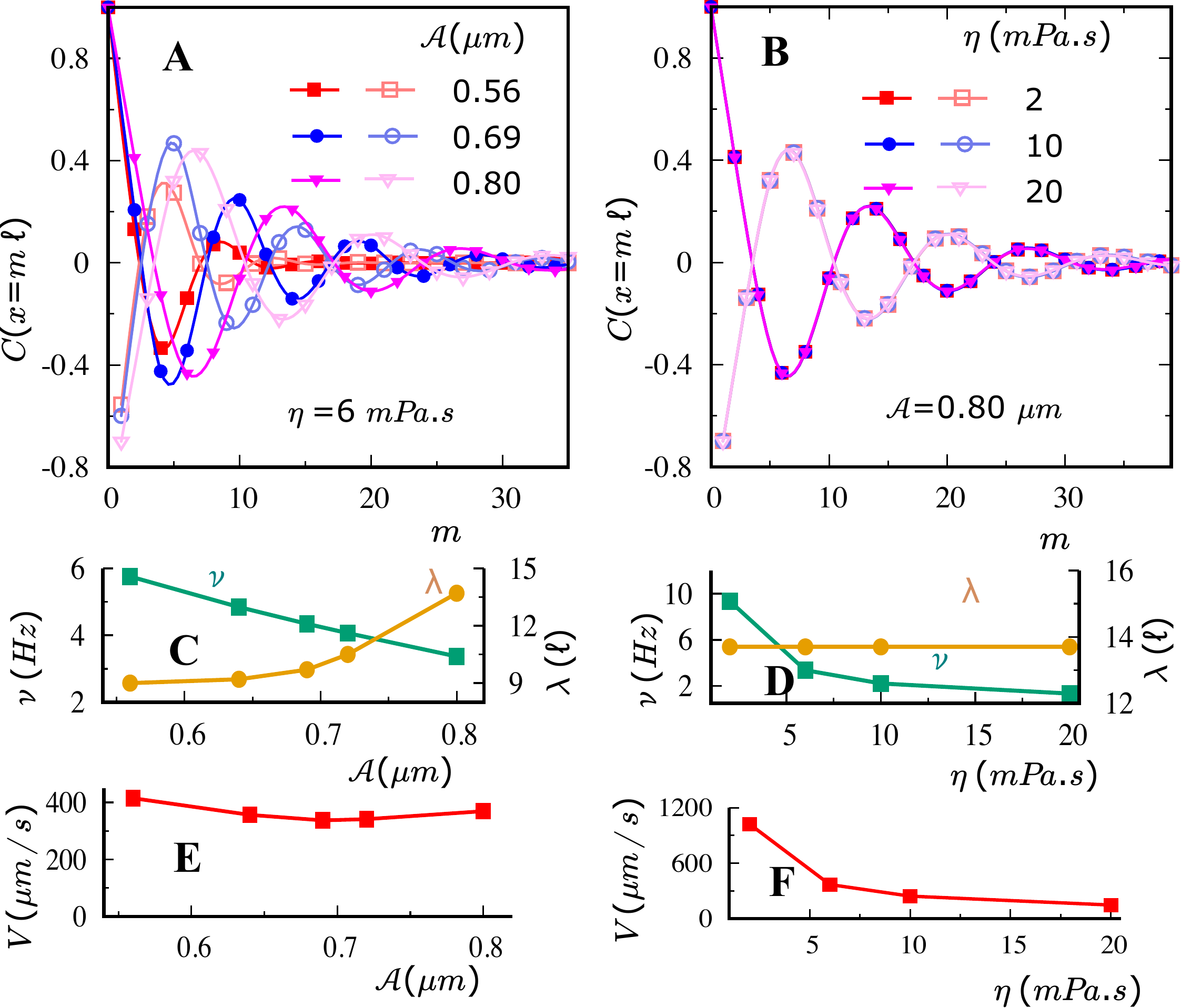}
\caption{Effect of viscosity and beating amplitude on metachronal waves in 1d ($N{=}200$). Correlation function $C(x{=}m\ell)$ as a function of the distance between rowers, for three different  $\mathcal{A}$ values (for $\eta{=}6\ mPa.s$) (A) and for three different $\eta$ values (for $\mathcal{A}{=}0.8\ \mu m$) (B). The wavelengths $\lambda$ are plotted against $\mathcal{A}$ (C), and $\eta$ (D), together with the corresponding beating frequency $\nu$. Propagation velocity $V$ is plotted against $\mathcal{A}$ (E), and $\eta$ (F). In (B), the different symbols are not distinctively visible as the three curves completely overlap.}
\label{fig:fig3}
\end{figure}

\subsection{A. Characterization of metachronal waves}
Figure.~\ref{fig:fig2} shows the metachronal waves on the one-dimensional lattice. The beads' displacement against the rowers' position for a given time displays two spatial waves that are visualized by connecting displacements $y_i$ by lines for all rowers at the even and odd lattice sites separately (Fig.~\ref{fig:fig2}A) in agreement with \cite{CosentinoPre03,StarkEpj11}. This is a unique feature of the rower model, and arises due to a degree of anti-phase synchronization between two adjacent rowers. 
The wave propagation is illustrated in Fig.~\ref{fig:fig2}C by the kymograph obtained for ${\sigma_i}(t)$.
To characterize it, we compute the spatial correlation function between two rowers as a function of their separation vector {$\bf r$}:
\bea
C({\bf r})=\frac{\sum_{i j} \la \sigma_i({\bf r}_i,t) \sigma_j ({\bf r}_j,t) \ra  \delta({\bf r}-{\bf r}_{ij})}
                            {\sum_{i j}\delta({\bf r}-{\bf r}_{ij})}.
\label{eqn:cr}                            
\eea
As the rowers are placed on a regular lattice with lattice spacing $\ell$, the coordinates of ${\bf r}$ are discrete and can be written as $\left(m\ell,n \ell\right)$ with $m,n\in\{0,1,2,..., L\}$. The measurement is done after a large equilibration time $t_0$, where the system is assumed to reach a steady state. Brackets $\langle \cdot \rangle$ represent average over  times and ensembles. An ensemble is the collection of 5000 sets of $\{\sigma_i(t)\}$ recorded every 2 seconds after $t_0$=2500 seconds. \textcolor{red}{Depending on system size, we consider several ensembles with random initial conditions. For 1d ($N{=}200$) the number of ensembles is $100$ and for 2d ($N{=}1600$),  it lies between $5$ to $10$.}

\subsection{B. Wave properties in one dimension}

Fig.~\ref{fig:fig2}B shows the variation of $C(x)=C(x, y=0)$ in one dimension. For odd and even $m$ values, two oscillating curves decay to zero as the distance between rowers $x=m\ell$ increases.
While the oscillations indicate the wave nature of the collective beating, the loss of correlations at larger $x$ suggest a damping in the coordination on a characteristic length scale $l_d$. $C(x)$ can be fitted with the simple function $\pm{{\rm e}^{-x/l_d} \cos({2 \pi x/ \lambda })}$, the $+$ ($-$) sign being for even (odd) $m$. This fit estimates the wavelength $\lambda$ and decay length $l_d$. For Fig.~\ref{fig:fig2}, $\lambda\simeq 13.7 \ell$ and $l_d\simeq9.0\ell$.
In a recent work, wavelength, and decay length were measured experimentally for metachronal waves on the human bronchial epithelium, and these two lengthscale values are comparable \cite{Mesdjian2022}. Our results are consistent with the experiment.
The ensemble and spatial average of the beating frequency was computed: $\nu\simeq 3.4 \ Hz$ and combined with $\lambda$ to infer the metachronal wave velocity $V=\nu\lambda\simeq 370\ \mu m.s^{-1}$. These values are consistent with estimates that can be inferred directly from the slopes in the kymograph Fig.~\ref{fig:fig2}C.
\begin{figure}[t]
\includegraphics[width=0.48\textwidth]{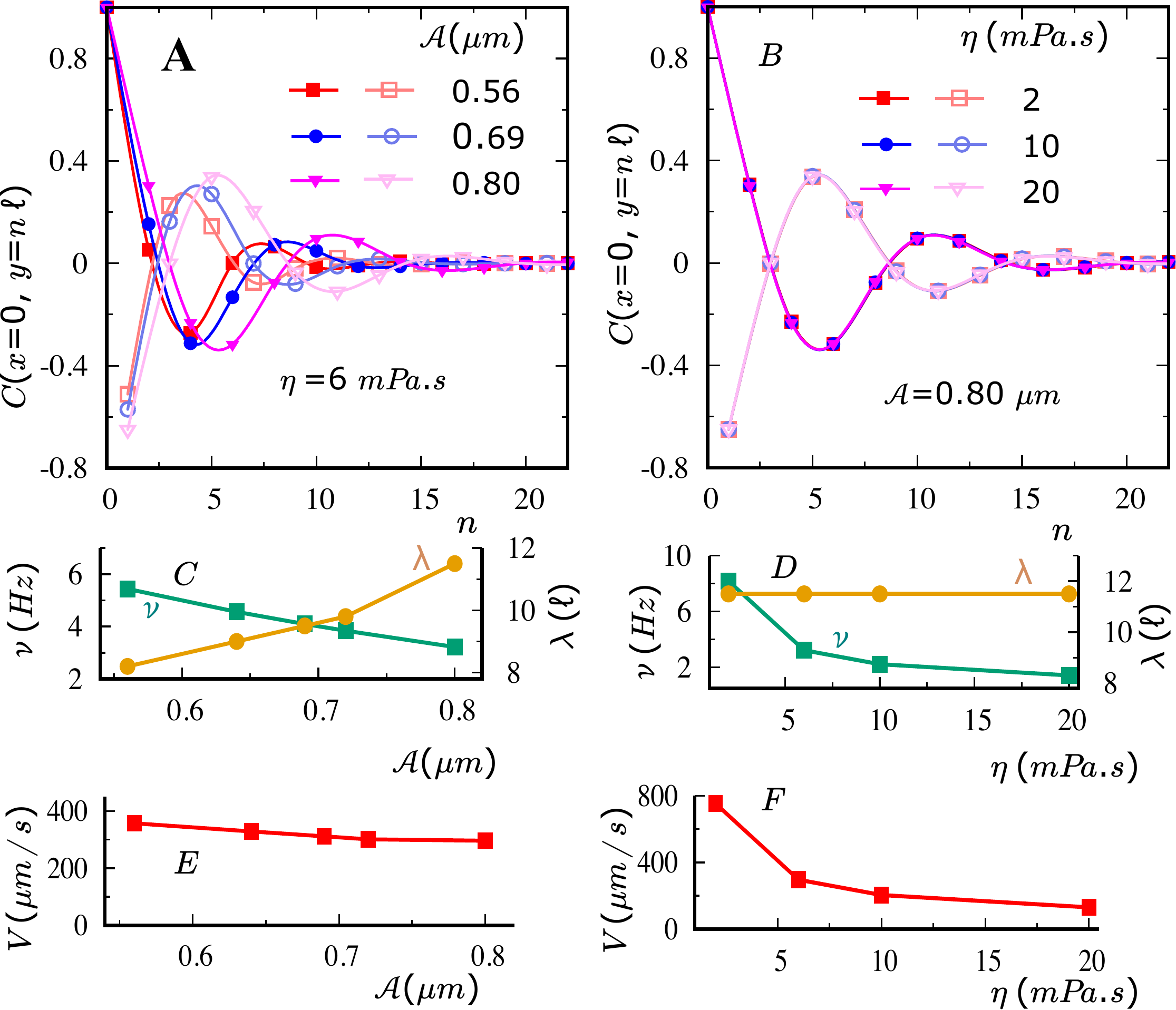}
\caption{Effect of viscosity and beating amplitude on metachronal waves in 2d ($N{=}L^2{=}1600$). Correlation function along beating direction $C(x{=}0, y{=}n\ell)$ as a function of the distance between rowers, for three different values of $\mathcal{A}$ (for $\eta{=}6\ mPa.s$) (A) and for three different $\eta$ values (for $\mathcal{A}{=}0.8\ \mu m$) (B). The wavelengths $\lambda$ are plotted against $\mathcal{A}$ (C), and $\eta$ (D), together with the corresponding beating frequency $\nu$. Propagation velocity $V$ is plotted as a function of $\mathcal{A}$ (E), and $\eta$ (F). In (B), the different symbols are not distinctively visible as the three curves completely overlap.}
\label{fig:fig4}
\end{figure}

We then investigate the effect of viscosity of the fluid medium and activity of the cilia on the metachronal waves quantities: $\lambda$, $l_d$, $\nu$, and $V{=}\nu\lambda$,  by computing  $C(x)$ for various $\eta$ and $\mathcal{A}$. The plot of $C(x)$ for different $\mathcal{A}$ values shows that both  $\lambda$ and  $l_d$ increase with $\mathcal{A}$ (Fig.~\ref{fig:fig3}A and C), whereas $\nu$ decreases with $\mathcal{A}$, keeping $V$ almost constant (Fig.~\ref{fig:fig3}C and E). As $\nu_{0}$, the frequency of a rower, decreases with $\mathcal{A}$, the decrease of $\nu$ is expected. On the same line, increasing $\mathcal{A}$, which is a characteristic length of the problem, may naturally increase the length scale of the emerging collective dynamics. Thus the respective variation of $\nu$ and $\lambda$ can be generally expected. What is remarkable though is that they compensate to result in an almost constant wave velocity.  
Interestingly, $C(x)$ does not depend on the values of $\eta$ (see Fig.~\ref{fig:fig3}B), meaning $\lambda$ and $l_d$ are independent of $\eta$ and implying that the spatial behavior of emergent waves does not depend on the fluid viscosity and are only determined by cilia activity parameters, in agreement with experimental observations \cite{Gheber1998,Machemer1972}.
%Below, we argued that such behavior is generally characteristic of a hydrodynamically coupled system. 
Finally, the frequency $\nu$ decreases as a function of $\eta$ and so does $V$, as measured experimentally in \cite{Gheber1998, Machemer1972}. 

\subsection{C. Wave properties in two dimensions}

\begin{figure}[t]
\includegraphics[width=0.48\textwidth]{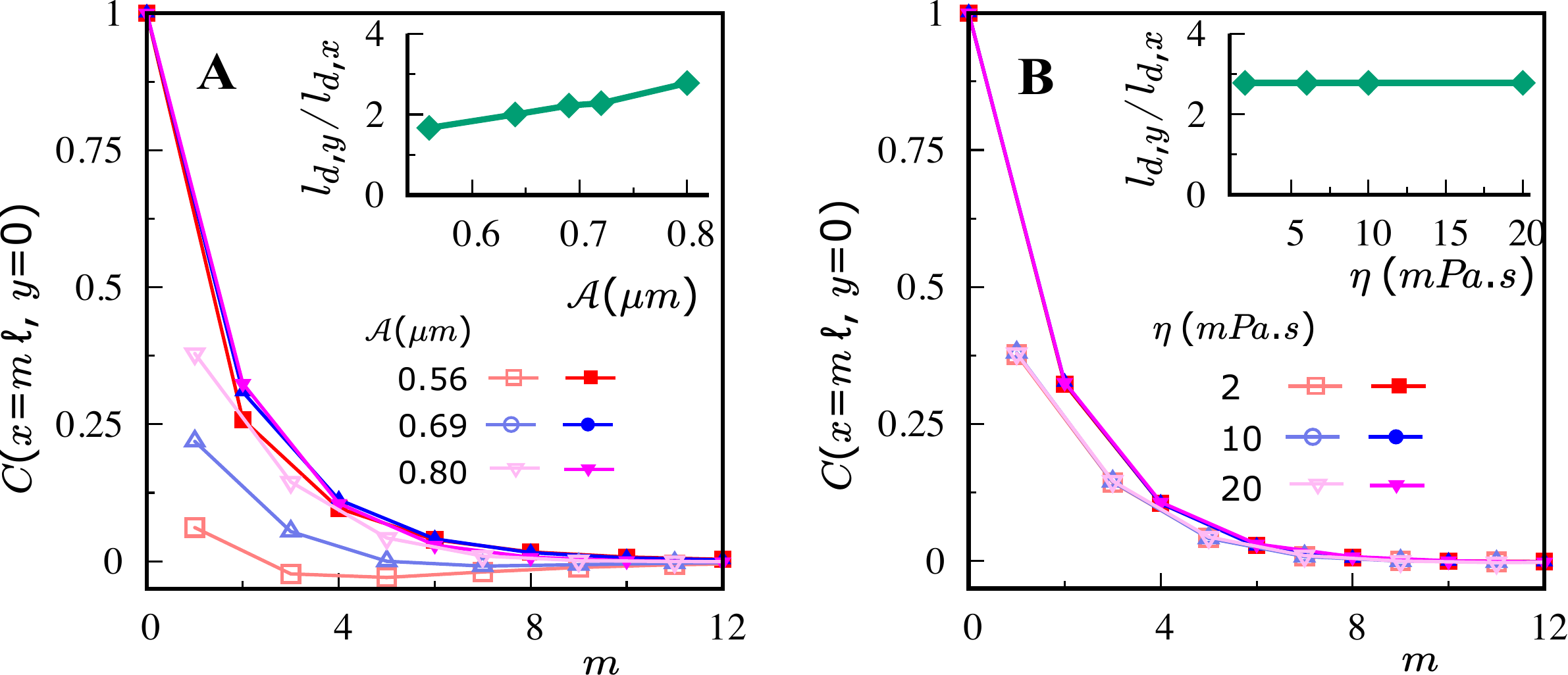}
\caption{Effect of viscosity and beating amplitude on correlations along the direction perpendicular to the beating direction $C(x{=}m\ell,y{=}0)$. A. $C(x{=}m\ell,y{=}0)$ for three different $\mathcal{A}$ for a fixed $\eta{=}6\, \rm{mPa{\cdot}s}$ (A), and for three values of $\eta$ for a given $\mathcal{A}{=}0.69 \mu m$ (B).}
\label{fig:fig5}
\end{figure}

On a square lattice, we find that the metachronal wave propagates along the beating direction $y$ whereas no wave is obtained in the perpendicular direction (Fig.~\ref{fig:fig5}), suggesting longitudinal waves.
%not $x$ direction, unlike in 1d. 
In Fig.~\ref{fig:fig4}A and B, we plot the correlation function along $y$-direction $C(0, y=n\ell)$ against $n$ for various values of $\mathcal{A}$ and $\eta$.
Similar to 1d, two spatial waves can be seen for even and odd values of $n$ for a given parameter set. 
For a fixed $\eta$, $\lambda$ increases and $\nu$ decreases with $\mathcal{A}$, keeping the wave velocity $V$ almost constant (Fig.~\ref{fig:fig4}C and E). On the contrary, $\lambda$ remains constant and $\nu$ decreases with $\eta$, leading to a decrease in $V$ with $\eta$ (Fig.~\ref{fig:fig4}D and F). 
%These results are consistent with the 1d results.
We further note that although the qualitative behavior of metachronal waves in 1d and 2d are similar, the values of $\lambda$ and $V$ are relatively larger in 1d. 
This result raises interesting questions on the implications of the geometry of realistic ciliated tissues, which are mostly organized in 2d groups of cilia bundles.

In the direction perpendicular to beating, no oscillation is obtained (Fig.~\ref{fig:fig5}).
$C(x,0)$ either monotonically decays to zero as for large $\mathcal{A}$ or shows a negative correlation for small $y=m\ell$ with odd $m$ that eventually approaches zero for large $m$. 
For a given $\mathcal{A}$, $C(x,0)$ does not depend on $\eta$ (Fig.~\ref{fig:fig5}B), although odd and even $m$ can follow different curves, reminiscent of $C(0,y)$. We compare the decay lengths of correlations along $x$ and $y$ directions  $l_{d,x}$ and $l_{d,y}$. The decay length  for the damped oscillations along $y$, $l_{d,y}$, can be estimated from the fitting method discussed above. The decay length  $l_{d,x}$ is estimated from the exponential fit of the $C(x=m\ell,0)$ for even $m$ values. The ratio $l_{d,y}/l_{d,x}$ is plotted in the insets, one notes that $l_{d,y}/l_{d,x} \gtrsim 2$. For a fixed $\mathcal{A}$, it remains unchanged with $\eta$. However, for a given $\eta$, the ratio increases with cilia activity $\mathcal{A}$, which implies an enhancement of coherence along the beating direction compared the perpendicular one. This anisotropic response may be related to the anisotropy of the interaction strength. Indeed, considering the same $r_{ij}$ value, $O(i,j)$ is two times larger along the $y$-axis than along the $x$-axis. 

\section{IV. Why spatial wave properties are independent of viscous media}

The fact that we obtain metachronal waves with spatial properties unaffected by viscosity has not been emphasized by previous studies, to our knowledge. Nevertheless, this remarkable numerical observation is robust on an order of magnitude of $\eta$ obtained with both 1d and 2d simulations. To rationalize this result, one needs to look into the details of characteristic length, and timescales of the system set by the activity and the surrounding viscous medium. The relaxation time $\tau_d = 6\pi\eta a/k$ for the bead motion in the viscous medium under a harmonic driving potential, which also determines the natural frequency $\nu_0$, is a crucial timescale in our problem. In the rower model, there are two lengths scales, $\mathcal{A}$, and $\mu$ (Fig.~\ref{fig:themodel}). Since we only vary $\mathcal{A}$, we chose it as the typical length scale. We note that our conclusion below, however, does not depend on the choice of the length scale. Multiplying both sides of Eq.~\ref{eqn:dyn_coupled} by $\tau_d / \mathcal{A}$ leads to an adimensional equation for the collective beating dynamics: 
\bea
\frac{d y_i^{\prime}}{d t^{\prime}} =  f^{\prime}_i + \sum_{i\neq j} \frac{3\,a}{4 r_{ij}} \left( 1+ (y_{ij}/r_{ij})^2 \right) f_j^{\prime} + \zeta_i(t^{\prime}),
\label{eqn:dimensionless}
\eea
where $t^{\prime}$ and $y^{\prime}$ are dimensionless time $t^{\prime}=t/\tau_d$ and displacement $y^{\prime}=y/\mathcal{A}$, $f^{\prime}_j = -(y_j^{\prime}-\mu \sigma_j/(2 \mathcal{A}))$ is the dimensionless force acting on rower $j$, and $\langle \zeta(t_1^{\prime}) \zeta(t_2^{\prime}) \rangle = 2 k_B T/(k \mathcal{A}^2) \delta(t^{\prime}_1 -t^{\prime}_2)$ is the adimensional noise correlation. As activity parameters $\mathcal{A}$, $k$, and $\mu$ are constant, Eq.~\ref{eqn:dimensionless} is $\eta$ independent. The latter means that the spatial properties are independent of $\eta$. 
However, as $\tau_d$ is affected by $\eta$,  it impacts the dynamical properties of the system. If any parameter $\mathcal{A}$, $k$, and $\mu$ are influenced by the medium, then our observation will break down.  

%We argue that the independence on fluid viscosity of the spatial emergent properties is more generic to systems operating at low Reynold's numbers, irrespective of the model details.For such systems, both the viscous drag and hydrodynamic coupling between two objects are inversely proportional to $\eta$. The thermal noise  $D$ and time $\tau_d$ are also inversely proportional to $\eta$. %As $\tau_d$ is inversely proportional to $\eta$, this also means that the normalization of Eq.~\ref{eqn:dyn_coupled} by $\tau_d$ will lead to the same conclusion. Therefore, one can get a similar adimensional equation as Eq.~\ref{eqn:dimensionless} for any model of active cilia coupled by a viscous fluid at low Reynolds numbers. Hence, the spatial properties of the emergent waves are expected to be independent of viscosity. If active beating is strongly dependent on the fluid rheology, then the active parameters of the rower model would depend on the viscosity, and this result would fail. Experimental results  \cite{Johnson1991} seem to indicate though a very small dependence of cilia beating amplitude with the liquid medium. We note that additional sources of deviation from this result could be the viscoelastic nature of the fluid or a non-thermal noise, which we have not addressed in this paper. Finally, although this simple dimensional analysis cannot predict the occurrence or nature of emergent behavior, it is powerful in predicting that the wavelength or other spatial properties will be viscosity independent in general.

\textcolor{red}{Above, we have studied the rower model, a minimal model for the active beating of cilia. The hydrodynamic interaction between cilia is implemented via Oseen coupling, assuming the beating amplitude is very small compared to intercellular distance and ignoring the surface effect and correlation in noise between two cilia. We also assumed that cilia are arranged in particular ways in space and beat in a specific direction. Here, we argue that the invariance of spatial properties of metachronal waves in various viscous media is more general and not dependent on the model chosen for cilia beating, nor on the cilia mutual interaction in the viscous fluid, nor on the cilia spatial arrangement. Indeed, for a more realistic cilia model, one may choose a complex force profile for active beating having more parameters ${\bf f}_i$ \cite{Guirao2007,Chakrabarti2022}. For the hydrodynamic coupling, the presence of the surface, can be captured by the Blake tensor in the far field limit \cite{blake1971note}. At low Reynolds number, the hydrodynamic coupling tensor have the following properties: it has a multiplicative scalar factor that is inversely proportional to the drag coefficient $\gamma$, and the tensorial part depends on the position of the active forces.}  

%In the Stokes (LOW RE...?) regime
\textcolor{red}{
For a generic hydrodynamic coupling $G({\bf r}_i,{\bf r}_j)$ and correlated noise between cilia, the equation of motion for the position of the active beating forces is given by:  
\begin{eqnarray}
\frac{d {\bf r}_i}{dt} =    \sum_{j=1}^{N}   G({\bf r}_i,{\bf r}_j) \left( {\bf f}_j + {\bf f}^r_j(t) \right),
\end{eqnarray}
where the random force (noise) ${\bf f}_i^r$ obeys $\langle {\bf f}^r_i(t_1) {\bf f}^r_j(t_2) \rangle = 2 k_B T G^{-1}({\bf r}_i,{\bf r}_j) \delta(t_1-t_2)$ \cite{Ermak1978,Dufresne2008,Hamilton2021}.  As we discussed above, the coupling tensor can be written as $G({\bf r}_i,{\bf r}_j) = \tilde{G} ({\bf r}_i,{\bf r}_j))/\gamma$, with $\tilde{G}$ being  dimensionless and independent of viscosity. We note that the latter may not hold true for the viscoelastic or other complex fluids, which is not scope of this paper.  As the movement of a cilium is localized around an fixed position, for realistic cilia, there may at least exist an effective force constant $k^{eff}$ and a beating amplitude $\mathcal{A}^{eff}$. Therefore relevant lengthscale and timescale could respectively be written as $\mathcal{A}^{eff}$ and $\tau_d=\gamma/k^{eff}$, respectively. The important assumption here is that these activity parameters $k^{eff}$, and $\mathcal{A}^{eff}$ are independent of fluid viscosity. Now, the dimensionless dynamical equation for cilia $i$ is given by:
\bea
\frac{d {\bf r}_i^{\prime}}{dt^{\prime}} =    \sum_{j=1}^{N}  \tilde{G}({\bf r}_i,{\bf r}_j)\left( {\bf f^{\prime}}_j + {\bf f}^{r,\prime}_j(t^{\prime}) \right)
\label{eqn:dimensionless_general}
\eea
where ${\bf f}^{\prime}_j{=}{\bf f}_j/(k^{eff} \mathcal{A}^{eff})$, is the dimensionless active force on cilia $j$. For the dimensionless random forces, it can be shown that they obey the following relation: $\langle {\bf f}^{r,\prime}_i(t_1^{\prime}) {\bf f}^{r,\prime}_j(t_2^{\prime}) \rangle {=} 2 k_B T/(k^{eff} ({\mathcal{A}}^{eff})^2)  \tilde{G}^{-1}({\bf r}_i,{\bf r}_j) \delta(t^{\prime}_1 -t^{\prime}_2)$. Therefore, for a general model for cilia with hydrodynamic coupling, the equation can be written in dimensionless form as in Eq.~\ref{eqn:dimensionless_general}. Thus, in the steady-state, the spatial properties of the emergent waves will be independent of viscosity, meaning that dissipation only affects temporal parameters.}

 %I WOULD SUPRESS THIS LAST SENTENCE: However, the quantitative behavior of metachronal waves can depend on the details of ciliary model and hydrodynamic interactions.  }

\section{V. Discussion and Conclusion}
\textcolor{red}{
 Although the dimensional analysis cannot predict the occurrence or nature of emergent behavior, it is remarkable in predicting that the wavelength or other spatial properties will be viscosity-independent in general. This property of hydrodynamically coupled oscillators could be used as a simple framework to understand the origin of metachronal waves in biological systems. The breakdown of such invariance can indicate the complexity of living systems. We discuss three following scenarios where our prediction regarding independence of spatial wave properties with viscosity may fail.
 }

\begin{itemize}[leftmargin=*]

     \item  \textcolor{red}{For real cilia, the  viscosity of the medium may influence the activity parameters, and then our observation can break down. As we discussed above, in this scenario the right hand side of the equation Eq.\ref{eqn:dimensionless}, will be dependent of viscosity. 
     Therefore, we may not expect invariance of spatial properties of emergent waves. Experimental results \cite{Johnson1991, Katoh2018} seem to indicate a small dependency of cilia beating amplitude with the liquid medium. Increased viscosity can also alter the beating orientation \cite{Machemer1972,Gheber1998}, amplify the asymmetry between stroke and anti-stroke \cite{Katoh2018}, and change other details of the beating pattern \cite{Brokaw1966}. These can cause changes in the metachronal wave characteristics and in particular in the direction of propagation. %In our simple model, this is not a result that could be obtained, mainly because all rowers beat in the same direction and because the symmetry is broken, the beating direction being in the plane of the rowers.
     These experimental results indicate that the beating machinery could adapt to the viscous load.}

%Increased viscosity can also alter the beating orientation \cite{Machemer1972,Gheber1998}, amplifies the asymmetry between stroke and anti-stroke \cite{Katoh2018}, and other details of the beating pattern \cite{Brokaw1966}.

%the assymmetry between stroke and anti-stroke is enhanced when the viscosity of the surrounding fluid increases.    
    % {\color{teal} Experimental results  \cite{Johnson1991} seem to indicate a small dependency of cilia beating amplitude with the liquid medium. In \cite{Machemer1972} and \cite{Gheber1998}, but also in [Goldstein RE, Lauga E, Pesci AI, Proctor MRE. 2016. Elastohydrodynamic synchronization of adjacent beating flagella. Physical Review Fluids 1:073201. DOI: https://doi.org/10.1103/PhysRevFluids.1.073201, PMID: 29750206], a waveform compliance is obtained, i.e. the assymmetry between stroke and anti-stroke is enhanced when the viscosity of the surrounding fluid increases.
     
     \item \textcolor{red}{The fluctuations in our study are thermal, where an effective fluctuation-dissipation theorem is expected to hold. However, some experiments point out that in real cilia and flagella, the fluctuations are mostly dominated by those that have an athermal origin \cite{Polin2009,Goldstein2009PRL,Ma2014,Wan2014PRL,Quaranta2015}. To understand how this athermal noise affects spatial properties of metachronal waves, we numerically studied the model in presence of noise with a strength kept constant against viscosity. We numerically solve Eq.~\ref{eqn:dyn_coupled} for different values of $\eta$, with a constant noise strength $D$. The results are plotted in Fig.~\ref{fig:fig6}. There is thus no thermal noise in this simulation and we expected spatial properties then to depend on the viscosity as Eq.~\ref{eqn:dimensionless}, will be dependent of viscosity. We indeed observe that the correlation length scale $l_d$ decays with viscosity, while the wavelength remains almost constant. This is in contrast with the previous (thermal) case where $l_d$ was constant as the viscosity was varied. It also means that the wavelength is not much influenced by dissipation.}

\begin{figure}[t]
\includegraphics[width=0.48\textwidth]{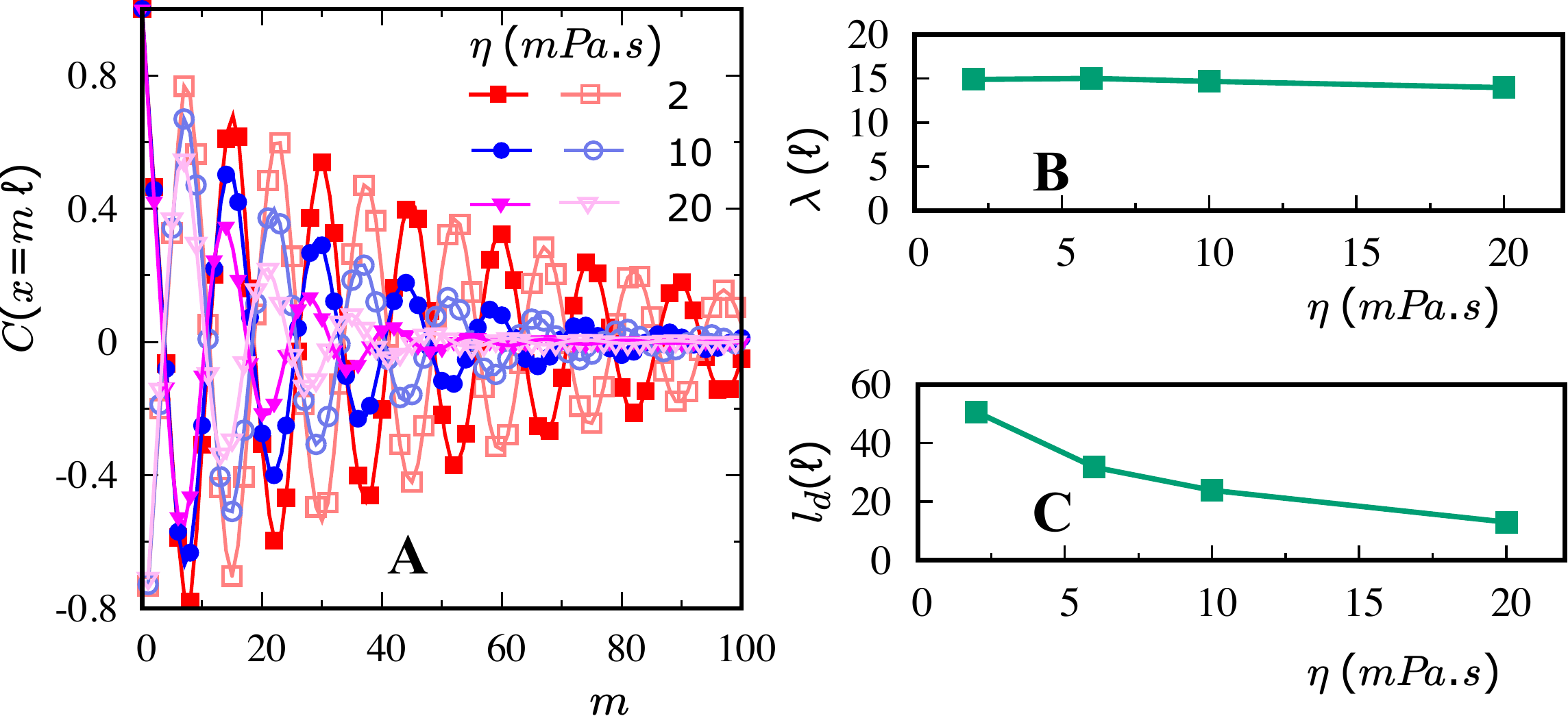}
\caption{Effect of viscosity in presence of athermal noise in 1d ($N{=}L{=}200$). Correlation function $C(x{=}m\ell)$ as a function of the distance between rowers for three different $\eta$ values (for $\mathcal{A}{=}0.8\ \mu m$) (B). The wavelength $\lambda$ (in units of $\ell$) is plotted against $\eta$. (C) Decay length $l_d$ (in units of $\ell$) is plotted against $\eta$. Unlike thermal noise where diffusivity is $D=k_BT/\gamma$, in this athermal case, the diffusivity $D$ is kept constant to a value (${=}0.01 \mu m^2s^{-1}$) while we vary $\eta$.}
\label{fig:fig6}
\end{figure}
 %\begin{itemize}[leftmargin=*]
 
     \item \textcolor{red}{Another source of non-invariance of spatial properties could be the viscoelastic nature of the fluid. For the viscoelastic fluids, the timescales could be nonlinear as a function of viscosity. The above dimensional argument will not hold, and we believe that the invariance of spatial properties will not be true in such a case. 
     Experimentally, the wavelength seems to be almost constant within a wide range of fluid viscosities \cite{Gheber1998,Machemer1972}, while the frequency $\nu$ and $V$ decrease as a function of $\eta$. We note that in \cite{Gheber1990}, the viscosity is varied on a very large range. For viscosity larger than $\rm{20\,mPa{\cdot}s}$, we suspect that the liquid is no more newtonian and probably viscoelastic given the high molecular weight polymers used for the viscous solutions. For these higher viscosities, another lower plateau in wavelength is found: the wavelength seems to show a sudden change after a critical viscosity value \cite{Gheber1990}, after which the wavelength reaches a lower value and remains constant.} 
     %We believe that this could be due to the sudden change in the beating angle of cilia after that critical viscosity.
  \end{itemize}
     
     \textcolor{red}{In a recent work by Ringers et al. \cite{Ringers2023}, an increase in spatial correlation length ($l_d$) is observed with viscosity, where methycellulose concentration is varied to increase medium viscosity. Why this coherence length scale increases with viscosity is not clear. This result cannot be explained by our model neither in the presence of athermal noise (Fig.~\ref{fig:fig6}) nor thermal noise. It is known that a higher methylcellulose concentration not only increases viscosity but also makes the fluid viscoelastic. Whether viscoelasticity plays role to increase the spatial coherence needs further investigations which is not scope of this paper. }

In conclusion, using a simple rower model of coupled oscillators, we have studied the influence of activity and dissipation on the spatial and temporal synchronization properties of cilia assemblies. Enhancement of cilia activity increases the wavelength and beating period, keeping the wave velocity almost unchanged. Remarkably, the viscosity does not affect spatial patterns of metachronal waves. Using dimensional analysis, we demonstrate that this result is robust against complexity of cilia model and hydrodynamic coupling due to viscous media. On the contrary, the beating frequency and the wave velocity indeed decrease with viscosity. The deviation from such a behaviour may indicate influence of medium on cilia activity, the presence of athermal fluctuations, or viscoelastic nature of the medium. Our findings could pave the way in understanding the emergence of specific functions of cilia in pathological contexts, for example \cite{Kanale2022}. 
 
%We argue that the independence on fluid viscosity of the spatial emergent properties is more generic to systems operating at low Reynold's numbers, irrespective of the model details. For such systems, both the viscous drag and hydrodynamic coupling between two objects are inversely proportional to $\eta$. The thermal noise  $D$ and time $\tau_d$ are also inversely proportional to $\eta$. %As $\tau_d$ is inversely proportional to $\eta$, this also means that the normalization of Eq.~\ref{eqn:dyn_coupled} by $\tau_d$ will lead to the same conclusion. Therefore, one can get a similar adimensional equation as Eq.~\ref{eqn:dimensionless} for any model of active cilia coupled by a viscous fluid at low Reynolds numbers. Hence, the spatial properties of the emergent waves are expected to be independent of viscosity. If active beating is strongly dependent on the fluid rheology, then the active parameters of the rower model would depend on the viscosity, and this result would fail. Experimental results  \cite{Johnson1991} seem to indicate though a very small dependence of cilia beating amplitude with the liquid medium. We note that additional sources of deviation from this result could be the viscoelastic nature of the fluid or a non-thermal noise, which we have not addressed in this paper. Finally, although this simple dimensional analysis cannot predict the occurrence or nature of emergent behavior, it is powerful in predicting that the wavelength or other spatial properties will be viscosity independent in general.

%\textcolor{magenta}{Add lacking references}

\bibliography{references_cilia}{}

\begin{thebibliography}{48}%
\makeatletter
\providecommand \@ifxundefined [1]{%
 \@ifx{#1\undefined}
}%
\providecommand \@ifnum [1]{%
 \ifnum #1\expandafter \@firstoftwo
 \else \expandafter \@secondoftwo
 \fi
}%
\providecommand \@ifx [1]{%
 \ifx #1\expandafter \@firstoftwo
 \else \expandafter \@secondoftwo
 \fi
}%
\providecommand \natexlab [1]{#1}%
\providecommand \enquote  [1]{``#1''}%
\providecommand \bibnamefont  [1]{#1}%
\providecommand \bibfnamefont [1]{#1}%
\providecommand \citenamefont [1]{#1}%
\providecommand \href@noop [0]{\@secondoftwo}%
\providecommand \href [0]{\begingroup \@sanitize@url \@href}%
\providecommand \@href[1]{\@@startlink{#1}\@@href}%
\providecommand \@@href[1]{\endgroup#1\@@endlink}%
\providecommand \@sanitize@url [0]{\catcode `\\12\catcode `\$12\catcode
  `\&12\catcode `\#12\catcode `\^12\catcode `\_12\catcode `\%12\relax}%
\providecommand \@@startlink[1]{}%
\providecommand \@@endlink[0]{}%
\providecommand \url  [0]{\begingroup\@sanitize@url \@url }%
\providecommand \@url [1]{\endgroup\@href {#1}{\urlprefix }}%
\providecommand \urlprefix  [0]{URL }%
\providecommand \Eprint [0]{\href }%
\providecommand \doibase [0]{http://dx.doi.org/}%
\providecommand \selectlanguage [0]{\@gobble}%
\providecommand \bibinfo  [0]{\@secondoftwo}%
\providecommand \bibfield  [0]{\@secondoftwo}%
\providecommand \translation [1]{[#1]}%
\providecommand \BibitemOpen [0]{}%
\providecommand \bibitemStop [0]{}%
\providecommand \bibitemNoStop [0]{.\EOS\space}%
\providecommand \EOS [0]{\spacefactor3000\relax}%
\providecommand \BibitemShut  [1]{\csname bibitem#1\endcsname}%
\let\auto@bib@innerbib\@empty
%</preamble>
\bibitem [{\citenamefont {Gray}(1928)}]{Gray1928}%
  \BibitemOpen
  \bibfield  {author} {\bibinfo {author} {\bibfnamefont {J.}~\bibnamefont
  {Gray}},\ }\href@noop {} {\emph {\bibinfo {title} {Ciliary movement}}}\
  (\bibinfo  {publisher} {Cambridge University Press},\ \bibinfo {year}
  {1928})\BibitemShut {NoStop}%
\bibitem [{\citenamefont {Sleigh}(1962)}]{Sleigh1962}%
  \BibitemOpen
  \bibfield  {author} {\bibinfo {author} {\bibfnamefont {M.~A.}\ \bibnamefont
  {Sleigh}},\ }\href@noop {} {\emph {\bibinfo {title} {The Biology of Cilia and
  Flagella}}}\ (\bibinfo  {publisher} {Pergamon Press},\ \bibinfo {year}
  {1962})\BibitemShut {NoStop}%
\bibitem [{\citenamefont {Blake}(1972)}]{Blake1972}%
  \BibitemOpen
  \bibfield  {author} {\bibinfo {author} {\bibfnamefont {J.}~\bibnamefont
  {Blake}},\ }\href@noop {} {\bibfield  {journal} {\bibinfo  {journal} {Journal
  of Fluid Mechanics}\ }\textbf {\bibinfo {volume} {55}},\ \bibinfo {pages} {1}
  (\bibinfo {year} {1972})}\BibitemShut {NoStop}%
\bibitem [{\citenamefont {Gilpin}\ \emph {et~al.}(2020)\citenamefont {Gilpin},
  \citenamefont {Bull},\ and\ \citenamefont {Prakash}}]{Gilpin2020}%
  \BibitemOpen
  \bibfield  {author} {\bibinfo {author} {\bibfnamefont {W.}~\bibnamefont
  {Gilpin}}, \bibinfo {author} {\bibfnamefont {M.~S.}\ \bibnamefont {Bull}}, \
  and\ \bibinfo {author} {\bibfnamefont {M.}~\bibnamefont {Prakash}},\ }\href
  {\doibase 10.1038/s42254-019-0129-0} {\bibfield  {journal} {\bibinfo
  {journal} {Nature Reviews Physics}\ }\textbf {\bibinfo {volume} {2}},\
  \bibinfo {pages} {74} (\bibinfo {year} {2020})}\BibitemShut {NoStop}%
\bibitem [{\citenamefont {Milana}\ \emph {et~al.}(2020)\citenamefont {Milana},
  \citenamefont {Zhang}, \citenamefont {Vetrano}, \citenamefont {Peerlinck},
  \citenamefont {Volder}, \citenamefont {Onck}, \citenamefont {Reynaerts},\
  and\ \citenamefont {Gorissen}}]{Milana2020}%
  \BibitemOpen
  \bibfield  {author} {\bibinfo {author} {\bibfnamefont {E.}~\bibnamefont
  {Milana}}, \bibinfo {author} {\bibfnamefont {R.}~\bibnamefont {Zhang}},
  \bibinfo {author} {\bibfnamefont {M.~R.}\ \bibnamefont {Vetrano}}, \bibinfo
  {author} {\bibfnamefont {S.}~\bibnamefont {Peerlinck}}, \bibinfo {author}
  {\bibfnamefont {M.~D.}\ \bibnamefont {Volder}}, \bibinfo {author}
  {\bibfnamefont {P.~R.}\ \bibnamefont {Onck}}, \bibinfo {author}
  {\bibfnamefont {D.}~\bibnamefont {Reynaerts}}, \ and\ \bibinfo {author}
  {\bibfnamefont {B.}~\bibnamefont {Gorissen}},\ }\href {\doibase
  10.1126/sciadv.abd2508} {\bibfield  {journal} {\bibinfo  {journal} {Science
  Advances}\ }\textbf {\bibinfo {volume} {6}},\ \bibinfo {pages} {eabd2508}
  (\bibinfo {year} {2020})}\BibitemShut {NoStop}%
\bibitem [{\citenamefont {Brumley}\ \emph {et~al.}(2015)\citenamefont
  {Brumley}, \citenamefont {Polin}, \citenamefont {Pedley},\ and\ \citenamefont
  {Goldstein}}]{Brumley2015}%
  \BibitemOpen
  \bibfield  {author} {\bibinfo {author} {\bibfnamefont {D.~R.}\ \bibnamefont
  {Brumley}}, \bibinfo {author} {\bibfnamefont {M.}~\bibnamefont {Polin}},
  \bibinfo {author} {\bibfnamefont {T.~J.}\ \bibnamefont {Pedley}}, \ and\
  \bibinfo {author} {\bibfnamefont {R.~E.}\ \bibnamefont {Goldstein}},\ }\href
  {\doibase 10.1098/rsif.2014.1358} {\bibfield  {journal} {\bibinfo  {journal}
  {Journal of The Royal Society Interface}\ }\textbf {\bibinfo {volume} {12}},\
  \bibinfo {pages} {20141358} (\bibinfo {year} {2015})}\BibitemShut {NoStop}%
\bibitem [{\citenamefont {Zhang}\ \emph {et~al.}(2021)\citenamefont {Zhang},
  \citenamefont {den Toonder},\ and\ \citenamefont {Onck}}]{Zhang2021}%
  \BibitemOpen
  \bibfield  {author} {\bibinfo {author} {\bibfnamefont {R.}~\bibnamefont
  {Zhang}}, \bibinfo {author} {\bibfnamefont {J.}~\bibnamefont {den Toonder}},
  \ and\ \bibinfo {author} {\bibfnamefont {P.~R.}\ \bibnamefont {Onck}},\
  }\href {\doibase 10.1063/5.0054929} {\bibfield  {journal} {\bibinfo
  {journal} {Physics of Fluids}\ }\textbf {\bibinfo {volume} {33}},\ \bibinfo
  {pages} {092009} (\bibinfo {year} {2021})}\BibitemShut {NoStop}%
\bibitem [{\citenamefont {Wanner}\ \emph {et~al.}(1996)\citenamefont {Wanner},
  \citenamefont {Salath{\'e}},\ and\ \citenamefont {O'Riordan}}]{Wanner1996}%
  \BibitemOpen
  \bibfield  {author} {\bibinfo {author} {\bibfnamefont {A.}~\bibnamefont
  {Wanner}}, \bibinfo {author} {\bibfnamefont {M.}~\bibnamefont {Salath{\'e}}},
  \ and\ \bibinfo {author} {\bibfnamefont {T.~G.}\ \bibnamefont {O'Riordan}},\
  }\href@noop {} {\bibfield  {journal} {\bibinfo  {journal} {American journal
  of respiratory and critical care medicine}\ }\textbf {\bibinfo {volume}
  {154}},\ \bibinfo {pages} {1868} (\bibinfo {year} {1996})}\BibitemShut
  {NoStop}%
\bibitem [{\citenamefont {Smith}\ \emph {et~al.}(2008)\citenamefont {Smith},
  \citenamefont {Gaffney},\ and\ \citenamefont {Blake}}]{SmithRes08}%
  \BibitemOpen
  \bibfield  {author} {\bibinfo {author} {\bibfnamefont {D.}~\bibnamefont
  {Smith}}, \bibinfo {author} {\bibfnamefont {E.}~\bibnamefont {Gaffney}}, \
  and\ \bibinfo {author} {\bibfnamefont {J.}~\bibnamefont {Blake}},\
  }\href@noop {} {\bibfield  {journal} {\bibinfo  {journal} {Respiratory
  physiology and Nurobiology}\ }\textbf {\bibinfo {volume} {163}} (\bibinfo
  {year} {2008})}\BibitemShut {NoStop}%
\bibitem [{\citenamefont {Lauga}(2020)}]{Lauga2020}%
  \BibitemOpen
  \bibfield  {author} {\bibinfo {author} {\bibfnamefont {E.}~\bibnamefont
  {Lauga}},\ }\href@noop {} {\emph {\bibinfo {title} {The fluid dynamics of
  cell motility}}},\ Vol.~\bibinfo {volume} {62}\ (\bibinfo  {publisher}
  {Cambridge University Press},\ \bibinfo {year} {2020})\BibitemShut {NoStop}%
\bibitem [{\citenamefont {Lagomarsino}\ \emph {et~al.}(2003)\citenamefont
  {Lagomarsino}, \citenamefont {Jona},\ and\ \citenamefont
  {Bassetti}}]{CosentinoPre03}%
  \BibitemOpen
  \bibfield  {author} {\bibinfo {author} {\bibfnamefont {M.~C.}\ \bibnamefont
  {Lagomarsino}}, \bibinfo {author} {\bibfnamefont {P.}~\bibnamefont {Jona}}, \
  and\ \bibinfo {author} {\bibfnamefont {B.}~\bibnamefont {Bassetti}},\
  }\href@noop {} {\bibfield  {journal} {\bibinfo  {journal} {Physical Review
  E}\ }\textbf {\bibinfo {volume} {68}},\ \bibinfo {pages} {021908} (\bibinfo
  {year} {2003})}\BibitemShut {NoStop}%
\bibitem [{\citenamefont {Elgeti}\ and\ \citenamefont
  {Gompper}(2013)}]{Elgeti2013}%
  \BibitemOpen
  \bibfield  {author} {\bibinfo {author} {\bibfnamefont {J.}~\bibnamefont
  {Elgeti}}\ and\ \bibinfo {author} {\bibfnamefont {G.}~\bibnamefont
  {Gompper}},\ }\href {\doibase 10.1073/pnas.1218869110} {\bibfield  {journal}
  {\bibinfo  {journal} {Proc. Natl. Acad. Sci. USA}\ }\textbf {\bibinfo
  {volume} {110}},\ \bibinfo {pages} {4470} (\bibinfo {year}
  {2013})}\BibitemShut {NoStop}%
\bibitem [{\citenamefont {Golestanian}\ \emph {et~al.}(2011)\citenamefont
  {Golestanian}, \citenamefont {Yeomans},\ and\ \citenamefont
  {Uchida}}]{GolestanianRev11}%
  \BibitemOpen
  \bibfield  {author} {\bibinfo {author} {\bibfnamefont {R.}~\bibnamefont
  {Golestanian}}, \bibinfo {author} {\bibfnamefont {J.~M.}\ \bibnamefont
  {Yeomans}}, \ and\ \bibinfo {author} {\bibfnamefont {N.}~\bibnamefont
  {Uchida}},\ }\href@noop {} {\bibfield  {journal} {\bibinfo  {journal} {Soft
  Matter}\ }\textbf {\bibinfo {volume} {7}},\ \bibinfo {pages} {3074} (\bibinfo
  {year} {2011})}\BibitemShut {NoStop}%
\bibitem [{\citenamefont {Uchida}\ \emph {et~al.}(2017)\citenamefont {Uchida},
  \citenamefont {Golestanian},\ and\ \citenamefont {Bennett}}]{Uchida2017}%
  \BibitemOpen
  \bibfield  {author} {\bibinfo {author} {\bibfnamefont {N.}~\bibnamefont
  {Uchida}}, \bibinfo {author} {\bibfnamefont {R.}~\bibnamefont {Golestanian}},
  \ and\ \bibinfo {author} {\bibfnamefont {R.~R.}\ \bibnamefont {Bennett}},\
  }\href {\doibase 10.7566/JPSJ.86.101007} {\bibfield  {journal} {\bibinfo
  {journal} {Journal of the Physical Society of Japan}\ }\textbf {\bibinfo
  {volume} {86}},\ \bibinfo {pages} {101007} (\bibinfo {year}
  {2017})}\BibitemShut {NoStop}%
\bibitem [{\citenamefont {Chakrabarti}\ \emph {et~al.}(2022)\citenamefont
  {Chakrabarti}, \citenamefont {Fürthauer},\ and\ \citenamefont
  {Shelley}}]{Chakrabarti2022}%
  \BibitemOpen
  \bibfield  {author} {\bibinfo {author} {\bibfnamefont {B.}~\bibnamefont
  {Chakrabarti}}, \bibinfo {author} {\bibfnamefont {S.}~\bibnamefont
  {Fürthauer}}, \ and\ \bibinfo {author} {\bibfnamefont {M.~J.}\ \bibnamefont
  {Shelley}},\ }\href {\doibase 10.1073/pnas.2113539119} {\bibfield  {journal}
  {\bibinfo  {journal} {Proceedings of the National Academy of Sciences}\
  }\textbf {\bibinfo {volume} {119}},\ \bibinfo {pages} {e2113539119} (\bibinfo
  {year} {2022})}\BibitemShut {NoStop}%
\bibitem [{\citenamefont {Niedermayer}\ \emph {et~al.}(2008)\citenamefont
  {Niedermayer}, \citenamefont {Eckhardt},\ and\ \citenamefont
  {Lenz}}]{Niedermayer2008}%
  \BibitemOpen
  \bibfield  {author} {\bibinfo {author} {\bibfnamefont {T.}~\bibnamefont
  {Niedermayer}}, \bibinfo {author} {\bibfnamefont {B.}~\bibnamefont
  {Eckhardt}}, \ and\ \bibinfo {author} {\bibfnamefont {P.}~\bibnamefont
  {Lenz}},\ }\href {\doibase 10.1063/1.2956984} {\bibfield  {journal} {\bibinfo
   {journal} {Chaos: An Interdisciplinary Journal of Nonlinear Science}\
  }\textbf {\bibinfo {volume} {18}},\ \bibinfo {pages} {037128} (\bibinfo
  {year} {2008})}\BibitemShut {NoStop}%
\bibitem [{\citenamefont {Solovev}\ and\ \citenamefont
  {Friedrich}(2022{\natexlab{a}})}]{Solovev2022}%
  \BibitemOpen
  \bibfield  {author} {\bibinfo {author} {\bibfnamefont {A.}~\bibnamefont
  {Solovev}}\ and\ \bibinfo {author} {\bibfnamefont {B.~M.}\ \bibnamefont
  {Friedrich}},\ }\href {\doibase 10.1088/1367-2630/ac2ae4} {\bibfield
  {journal} {\bibinfo  {journal} {New Journal of Physics}\ }\textbf {\bibinfo
  {volume} {24}},\ \bibinfo {pages} {013015} (\bibinfo {year}
  {2022}{\natexlab{a}})}\BibitemShut {NoStop}%
\bibitem [{\citenamefont {Leoni}\ and\ \citenamefont
  {Liverpool}(2012)}]{LeoniPRE2012}%
  \BibitemOpen
  \bibfield  {author} {\bibinfo {author} {\bibfnamefont {M.}~\bibnamefont
  {Leoni}}\ and\ \bibinfo {author} {\bibfnamefont {T.~B.}\ \bibnamefont
  {Liverpool}},\ }\href {\doibase 10.1103/PhysRevE.85.040901} {\bibfield
  {journal} {\bibinfo  {journal} {Phys. Rev. E}\ }\textbf {\bibinfo {volume}
  {85}},\ \bibinfo {pages} {040901} (\bibinfo {year} {2012})}\BibitemShut
  {NoStop}%
\bibitem [{\citenamefont {Guirao}\ and\ \citenamefont
  {Joanny}(2007)}]{Guirao2007}%
  \BibitemOpen
  \bibfield  {author} {\bibinfo {author} {\bibfnamefont {B.}~\bibnamefont
  {Guirao}}\ and\ \bibinfo {author} {\bibfnamefont {J.~F.}\ \bibnamefont
  {Joanny}},\ }\href {\doibase 10.1529/biophysj.106.084897} {\bibfield
  {journal} {\bibinfo  {journal} {Biophysical Journal}\ }\textbf {\bibinfo
  {volume} {92}},\ \bibinfo {pages} {1900} (\bibinfo {year}
  {2007})}\BibitemShut {NoStop}%
\bibitem [{\citenamefont {Uchida}\ and\ \citenamefont
  {Golestanian}(2011)}]{Uchida2011}%
  \BibitemOpen
  \bibfield  {author} {\bibinfo {author} {\bibfnamefont {N.}~\bibnamefont
  {Uchida}}\ and\ \bibinfo {author} {\bibfnamefont {R.}~\bibnamefont
  {Golestanian}},\ }\href {\doibase 10.1103/PhysRevLett.106.058104} {\bibfield
  {journal} {\bibinfo  {journal} {Phys. Rev. Lett.}\ }\textbf {\bibinfo
  {volume} {106}},\ \bibinfo {pages} {058104} (\bibinfo {year}
  {2011})}\BibitemShut {NoStop}%
\bibitem [{\citenamefont {Brumley}\ \emph {et~al.}(2016)\citenamefont
  {Brumley}, \citenamefont {Bruot}, \citenamefont {Kotar}, \citenamefont
  {Goldstein}, \citenamefont {Cicuta},\ and\ \citenamefont
  {Polin}}]{Brumley2016}%
  \BibitemOpen
  \bibfield  {author} {\bibinfo {author} {\bibfnamefont {D.~R.}\ \bibnamefont
  {Brumley}}, \bibinfo {author} {\bibfnamefont {N.}~\bibnamefont {Bruot}},
  \bibinfo {author} {\bibfnamefont {J.}~\bibnamefont {Kotar}}, \bibinfo
  {author} {\bibfnamefont {R.~E.}\ \bibnamefont {Goldstein}}, \bibinfo {author}
  {\bibfnamefont {P.}~\bibnamefont {Cicuta}}, \ and\ \bibinfo {author}
  {\bibfnamefont {M.}~\bibnamefont {Polin}},\ }\href {\doibase
  10.1103/PhysRevFluids.1.081201} {\bibfield  {journal} {\bibinfo  {journal}
  {Phys. Rev. Fluids}\ }\textbf {\bibinfo {volume} {1}},\ \bibinfo {pages}
  {081201} (\bibinfo {year} {2016})}\BibitemShut {NoStop}%
\bibitem [{\citenamefont {Meng}\ \emph {et~al.}(2021)\citenamefont {Meng},
  \citenamefont {Bennett}, \citenamefont {Uchida},\ and\ \citenamefont
  {Golestanian}}]{Meng2021}%
  \BibitemOpen
  \bibfield  {author} {\bibinfo {author} {\bibfnamefont {F.}~\bibnamefont
  {Meng}}, \bibinfo {author} {\bibfnamefont {R.~R.}\ \bibnamefont {Bennett}},
  \bibinfo {author} {\bibfnamefont {N.}~\bibnamefont {Uchida}}, \ and\ \bibinfo
  {author} {\bibfnamefont {R.}~\bibnamefont {Golestanian}},\ }\href {\doibase
  10.1073/pnas.2102828118} {\bibfield  {journal} {\bibinfo  {journal}
  {Proceedings of the National Academy of Sciences}\ }\textbf {\bibinfo
  {volume} {118}},\ \bibinfo {pages} {e2102828118} (\bibinfo {year}
  {2021})}\BibitemShut {NoStop}%
\bibitem [{\citenamefont {Kanale}\ \emph {et~al.}(2022)\citenamefont {Kanale},
  \citenamefont {Ling}, \citenamefont {Guo}, \citenamefont {Fürthauer},\ and\
  \citenamefont {Kanso}}]{Kanale2022}%
  \BibitemOpen
  \bibfield  {author} {\bibinfo {author} {\bibfnamefont {A.~V.}\ \bibnamefont
  {Kanale}}, \bibinfo {author} {\bibfnamefont {F.}~\bibnamefont {Ling}},
  \bibinfo {author} {\bibfnamefont {H.}~\bibnamefont {Guo}}, \bibinfo {author}
  {\bibfnamefont {S.}~\bibnamefont {Fürthauer}}, \ and\ \bibinfo {author}
  {\bibfnamefont {E.}~\bibnamefont {Kanso}},\ }\href {\doibase
  10.1073/pnas.2214413119} {\bibfield  {journal} {\bibinfo  {journal}
  {Proceedings of the National Academy of Sciences}\ }\textbf {\bibinfo
  {volume} {119}},\ \bibinfo {pages} {e2214413119} (\bibinfo {year}
  {2022})}\BibitemShut {NoStop}%
\bibitem [{\citenamefont {Kotar}\ \emph {et~al.}(2010)\citenamefont {Kotar},
  \citenamefont {Leoni}, \citenamefont {Bassetti}, \citenamefont
  {Lagomarsino},\ and\ \citenamefont {Cicuta}}]{CicutaPnas10}%
  \BibitemOpen
  \bibfield  {author} {\bibinfo {author} {\bibfnamefont {J.}~\bibnamefont
  {Kotar}}, \bibinfo {author} {\bibfnamefont {M.}~\bibnamefont {Leoni}},
  \bibinfo {author} {\bibfnamefont {B.}~\bibnamefont {Bassetti}}, \bibinfo
  {author} {\bibfnamefont {M.~C.}\ \bibnamefont {Lagomarsino}}, \ and\ \bibinfo
  {author} {\bibfnamefont {P.}~\bibnamefont {Cicuta}},\ }\href@noop {}
  {\bibfield  {journal} {\bibinfo  {journal} {Proceedings of the National
  Academy of Sciences}\ }\textbf {\bibinfo {volume} {107}},\ \bibinfo {pages}
  {7669} (\bibinfo {year} {2010})}\BibitemShut {NoStop}%
\bibitem [{\citenamefont {Hamilton}\ and\ \citenamefont
  {Cicuta}(2021)}]{Hamilton2021}%
  \BibitemOpen
  \bibfield  {author} {\bibinfo {author} {\bibfnamefont {E.}~\bibnamefont
  {Hamilton}}\ and\ \bibinfo {author} {\bibfnamefont {P.}~\bibnamefont
  {Cicuta}},\ }\href {\doibase 10.1371/journal.pone.0249060} {\bibfield
  {journal} {\bibinfo  {journal} {PLOS ONE}\ }\textbf {\bibinfo {volume}
  {16}},\ \bibinfo {pages} {1} (\bibinfo {year} {2021})}\BibitemShut {NoStop}%
\bibitem [{\citenamefont {Solovev}\ and\ \citenamefont
  {Friedrich}(2022{\natexlab{b}})}]{Solovev2022b}%
  \BibitemOpen
  \bibfield  {author} {\bibinfo {author} {\bibfnamefont {A.}~\bibnamefont
  {Solovev}}\ and\ \bibinfo {author} {\bibfnamefont {B.~M.}\ \bibnamefont
  {Friedrich}},\ }\href {\doibase 10.1063/5.0075095} {\bibfield  {journal}
  {\bibinfo  {journal} {Chaos: An Interdisciplinary Journal of Nonlinear
  Science}\ }\textbf {\bibinfo {volume} {32}},\ \bibinfo {pages} {013124}
  (\bibinfo {year} {2022}{\natexlab{b}})},\ \Eprint
  {http://arxiv.org/abs/https://pubs.aip.org/aip/cha/article-pdf/doi/10.1063/5.0075095/16452347/013124\_1\_online.pdf}
  {https://pubs.aip.org/aip/cha/article-pdf/doi/10.1063/5.0075095/16452347/013124\_1\_online.pdf}
  \BibitemShut {NoStop}%
\bibitem [{\citenamefont {Dey}\ \emph {et~al.}(2018)\citenamefont {Dey},
  \citenamefont {Massiera},\ and\ \citenamefont {Pitard}}]{Deypre2018}%
  \BibitemOpen
  \bibfield  {author} {\bibinfo {author} {\bibfnamefont {S.}~\bibnamefont
  {Dey}}, \bibinfo {author} {\bibfnamefont {G.}~\bibnamefont {Massiera}}, \
  and\ \bibinfo {author} {\bibfnamefont {E.}~\bibnamefont {Pitard}},\ }\href
  {\doibase 10.1103/PhysRevE.97.012403} {\bibfield  {journal} {\bibinfo
  {journal} {Phys. Rev. E}\ }\textbf {\bibinfo {volume} {97}},\ \bibinfo
  {pages} {012403} (\bibinfo {year} {2018})}\BibitemShut {NoStop}%
\bibitem [{\citenamefont {Ramirez-San~Juan}\ \emph {et~al.}(2020)\citenamefont
  {Ramirez-San~Juan}, \citenamefont {Mathijssen}, \citenamefont {He},
  \citenamefont {Jan}, \citenamefont {Marshall},\ and\ \citenamefont
  {Prakash}}]{Ramirez2020}%
  \BibitemOpen
  \bibfield  {author} {\bibinfo {author} {\bibfnamefont {G.~R.}\ \bibnamefont
  {Ramirez-San~Juan}}, \bibinfo {author} {\bibfnamefont {A.~J.}\ \bibnamefont
  {Mathijssen}}, \bibinfo {author} {\bibfnamefont {M.}~\bibnamefont {He}},
  \bibinfo {author} {\bibfnamefont {L.}~\bibnamefont {Jan}}, \bibinfo {author}
  {\bibfnamefont {W.}~\bibnamefont {Marshall}}, \ and\ \bibinfo {author}
  {\bibfnamefont {M.}~\bibnamefont {Prakash}},\ }\href@noop {} {\bibfield
  {journal} {\bibinfo  {journal} {Nature physics}\ }\textbf {\bibinfo {volume}
  {16}},\ \bibinfo {pages} {958} (\bibinfo {year} {2020})}\BibitemShut
  {NoStop}%
\bibitem [{\citenamefont {Pellicciotta}\ \emph {et~al.}(2020)\citenamefont
  {Pellicciotta}, \citenamefont {Hamilton}, \citenamefont {Kotar},
  \citenamefont {Faucourt}, \citenamefont {Delgehyr}, \citenamefont {Spassky},\
  and\ \citenamefont {Cicuta}}]{Pellicciotta2020}%
  \BibitemOpen
  \bibfield  {author} {\bibinfo {author} {\bibfnamefont {N.}~\bibnamefont
  {Pellicciotta}}, \bibinfo {author} {\bibfnamefont {E.}~\bibnamefont
  {Hamilton}}, \bibinfo {author} {\bibfnamefont {J.}~\bibnamefont {Kotar}},
  \bibinfo {author} {\bibfnamefont {M.}~\bibnamefont {Faucourt}}, \bibinfo
  {author} {\bibfnamefont {N.}~\bibnamefont {Delgehyr}}, \bibinfo {author}
  {\bibfnamefont {N.}~\bibnamefont {Spassky}}, \ and\ \bibinfo {author}
  {\bibfnamefont {P.}~\bibnamefont {Cicuta}},\ }\href {\doibase
  10.1073/pnas.1910065117} {\bibfield  {journal} {\bibinfo  {journal}
  {Proceedings of the National Academy of Sciences}\ }\textbf {\bibinfo
  {volume} {117}},\ \bibinfo {pages} {8315} (\bibinfo {year}
  {2020})}\BibitemShut {NoStop}%
\bibitem [{\citenamefont {Machemer}(1972)}]{Machemer1972}%
  \BibitemOpen
  \bibfield  {author} {\bibinfo {author} {\bibfnamefont {H.}~\bibnamefont
  {Machemer}},\ }\href@noop {} {\bibfield  {journal} {\bibinfo  {journal} {J.
  Exp. Biol}\ ,\ \bibinfo {pages} {57}} (\bibinfo {year} {1972})}\BibitemShut
  {NoStop}%
\bibitem [{\citenamefont {Johnson}\ \emph {et~al.}(1991)\citenamefont
  {Johnson}, \citenamefont {Villalon}, \citenamefont {Royce}, \citenamefont
  {Hard},\ and\ \citenamefont {Verdugo}}]{Johnson1991}%
  \BibitemOpen
  \bibfield  {author} {\bibinfo {author} {\bibfnamefont {N.~T.}\ \bibnamefont
  {Johnson}}, \bibinfo {author} {\bibfnamefont {M.}~\bibnamefont {Villalon}},
  \bibinfo {author} {\bibfnamefont {F.~H.}\ \bibnamefont {Royce}}, \bibinfo
  {author} {\bibfnamefont {R.}~\bibnamefont {Hard}}, \ and\ \bibinfo {author}
  {\bibfnamefont {P.}~\bibnamefont {Verdugo}},\ }\href@noop {} {\bibfield
  {journal} {\bibinfo  {journal} {AM REV RESPIR DIS}\ }\textbf {\bibinfo
  {volume} {144}},\ \bibinfo {pages} {1091} (\bibinfo {year}
  {1991})}\BibitemShut {NoStop}%
\bibitem [{\citenamefont {Kikuchi}\ \emph {et~al.}(2017)\citenamefont
  {Kikuchi}, \citenamefont {Haga}, \citenamefont {Numayama-Tsuruta},
  \citenamefont {Ueno},\ and\ \citenamefont {Ishikawa}}]{Kikuchi2017}%
  \BibitemOpen
  \bibfield  {author} {\bibinfo {author} {\bibfnamefont {K.}~\bibnamefont
  {Kikuchi}}, \bibinfo {author} {\bibfnamefont {T.}~\bibnamefont {Haga}},
  \bibinfo {author} {\bibfnamefont {K.}~\bibnamefont {Numayama-Tsuruta}},
  \bibinfo {author} {\bibfnamefont {H.}~\bibnamefont {Ueno}}, \ and\ \bibinfo
  {author} {\bibfnamefont {T.}~\bibnamefont {Ishikawa}},\ }\href@noop {}
  {\bibfield  {journal} {\bibinfo  {journal} {Annals of biomedical
  engineering}\ }\textbf {\bibinfo {volume} {45}},\ \bibinfo {pages} {1048}
  (\bibinfo {year} {2017})}\BibitemShut {NoStop}%
\bibitem [{\citenamefont {Gheber}\ \emph {et~al.}(1998)\citenamefont {Gheber},
  \citenamefont {Korngreen},\ and\ \citenamefont {Priel}}]{Gheber1998}%
  \BibitemOpen
  \bibfield  {author} {\bibinfo {author} {\bibfnamefont {L.}~\bibnamefont
  {Gheber}}, \bibinfo {author} {\bibfnamefont {A.}~\bibnamefont {Korngreen}}, \
  and\ \bibinfo {author} {\bibfnamefont {Z.}~\bibnamefont {Priel}},\ }\href
  {\doibase
  https://doi.org/10.1002/(SICI)1097-0169(1998)39:1<9::AID-CM2>3.0.CO;2-3}
  {\bibfield  {journal} {\bibinfo  {journal} {Cell Motility}\ }\textbf
  {\bibinfo {volume} {39}},\ \bibinfo {pages} {9} (\bibinfo {year}
  {1998})}\BibitemShut {NoStop}%
\bibitem [{\citenamefont {Gheber}\ and\ \citenamefont
  {Priel}(1990)}]{Gheber1990}%
  \BibitemOpen
  \bibfield  {author} {\bibinfo {author} {\bibfnamefont {L.}~\bibnamefont
  {Gheber}}\ and\ \bibinfo {author} {\bibfnamefont {Z.}~\bibnamefont {Priel}},\
  }\href {\doibase https://doi.org/10.1002/cm.970160304} {\bibfield  {journal}
  {\bibinfo  {journal} {Cell Motility}\ }\textbf {\bibinfo {volume} {16}},\
  \bibinfo {pages} {167} (\bibinfo {year} {1990})}\BibitemShut {NoStop}%
\bibitem [{\citenamefont {Brumley}\ \emph {et~al.}(2012)\citenamefont
  {Brumley}, \citenamefont {Polin}, \citenamefont {Pedley},\ and\ \citenamefont
  {Goldstein}}]{Brumley2012}%
  \BibitemOpen
  \bibfield  {author} {\bibinfo {author} {\bibfnamefont {D.~R.}\ \bibnamefont
  {Brumley}}, \bibinfo {author} {\bibfnamefont {M.}~\bibnamefont {Polin}},
  \bibinfo {author} {\bibfnamefont {T.~J.}\ \bibnamefont {Pedley}}, \ and\
  \bibinfo {author} {\bibfnamefont {R.~E.}\ \bibnamefont {Goldstein}},\ }\href
  {\doibase 10.1103/PhysRevLett.109.268102} {\bibfield  {journal} {\bibinfo
  {journal} {Phys. Rev. Lett.}\ }\textbf {\bibinfo {volume} {109}},\ \bibinfo
  {pages} {268102} (\bibinfo {year} {2012})}\BibitemShut {NoStop}%
\bibitem [{\citenamefont {Wollin}\ and\ \citenamefont
  {Stark}(2011)}]{StarkEpj11}%
  \BibitemOpen
  \bibfield  {author} {\bibinfo {author} {\bibfnamefont {C.}~\bibnamefont
  {Wollin}}\ and\ \bibinfo {author} {\bibfnamefont {H.}~\bibnamefont {Stark}},\
  }\href@noop {} {\bibfield  {journal} {\bibinfo  {journal} {Eur. Phys. J. E}\
  }\textbf {\bibinfo {volume} {34}},\ \bibinfo {pages} {42} (\bibinfo {year}
  {2011})}\BibitemShut {NoStop}%
\bibitem [{\citenamefont {Mesdjian}\ \emph {et~al.}(2022)\citenamefont
  {Mesdjian}, \citenamefont {Wang}, \citenamefont {Gsell}, \citenamefont
  {D'Ortona}, \citenamefont {Favier}, \citenamefont {Viallat},\ and\
  \citenamefont {Loiseau}}]{Mesdjian2022}%
  \BibitemOpen
  \bibfield  {author} {\bibinfo {author} {\bibfnamefont {O.}~\bibnamefont
  {Mesdjian}}, \bibinfo {author} {\bibfnamefont {C.}~\bibnamefont {Wang}},
  \bibinfo {author} {\bibfnamefont {S.}~\bibnamefont {Gsell}}, \bibinfo
  {author} {\bibfnamefont {U.}~\bibnamefont {D'Ortona}}, \bibinfo {author}
  {\bibfnamefont {J.}~\bibnamefont {Favier}}, \bibinfo {author} {\bibfnamefont
  {A.}~\bibnamefont {Viallat}}, \ and\ \bibinfo {author} {\bibfnamefont
  {E.}~\bibnamefont {Loiseau}},\ }\href {\doibase
  10.1103/PhysRevLett.129.038101} {\bibfield  {journal} {\bibinfo  {journal}
  {Phys. Rev. Lett.}\ }\textbf {\bibinfo {volume} {129}},\ \bibinfo {pages}
  {038101} (\bibinfo {year} {2022})}\BibitemShut {NoStop}%
\bibitem [{\citenamefont {Blake}(1971)}]{blake1971note}%
  \BibitemOpen
  \bibfield  {author} {\bibinfo {author} {\bibfnamefont {J.~R.}\ \bibnamefont
  {Blake}},\ }in\ \href@noop {} {\emph {\bibinfo {booktitle} {Mathematical
  Proceedings of the Cambridge Philosophical Society}}},\ Vol.~\bibinfo
  {volume} {70}\ (\bibinfo {organization} {Cambridge University Press},\
  \bibinfo {year} {1971})\ pp.\ \bibinfo {pages} {303--310}\BibitemShut
  {NoStop}%
\bibitem [{\citenamefont {Ermak}\ and\ \citenamefont
  {McCammon}(1978)}]{Ermak1978}%
  \BibitemOpen
  \bibfield  {author} {\bibinfo {author} {\bibfnamefont {D.~L.}\ \bibnamefont
  {Ermak}}\ and\ \bibinfo {author} {\bibfnamefont {J.~A.}\ \bibnamefont
  {McCammon}},\ }\href {\doibase 10.1063/1.436761} {\bibfield  {journal}
  {\bibinfo  {journal} {The Journal of Chemical Physics}\ }\textbf {\bibinfo
  {volume} {69}},\ \bibinfo {pages} {1352} (\bibinfo {year}
  {1978})}\BibitemShut {NoStop}%
\bibitem [{\citenamefont {Dufresne}\ \emph {et~al.}(2000)\citenamefont
  {Dufresne}, \citenamefont {Squires}, \citenamefont {Brenner},\ and\
  \citenamefont {Grier}}]{Dufresne2008}%
  \BibitemOpen
  \bibfield  {author} {\bibinfo {author} {\bibfnamefont {E.~R.}\ \bibnamefont
  {Dufresne}}, \bibinfo {author} {\bibfnamefont {T.~M.}\ \bibnamefont
  {Squires}}, \bibinfo {author} {\bibfnamefont {M.~P.}\ \bibnamefont
  {Brenner}}, \ and\ \bibinfo {author} {\bibfnamefont {D.~G.}\ \bibnamefont
  {Grier}},\ }\href {\doibase 10.1103/PhysRevLett.85.3317} {\bibfield
  {journal} {\bibinfo  {journal} {Phys. Rev. Lett.}\ }\textbf {\bibinfo
  {volume} {85}},\ \bibinfo {pages} {3317} (\bibinfo {year}
  {2000})}\BibitemShut {NoStop}%
\bibitem [{\citenamefont {Katoh}\ \emph {et~al.}(2018)\citenamefont {Katoh},
  \citenamefont {Ikegami}, \citenamefont {Uchida}, \citenamefont {Iwase},
  \citenamefont {Nakane}, \citenamefont {Masaike}, \citenamefont {Setou},\ and\
  \citenamefont {Nishizaka}}]{Katoh2018}%
  \BibitemOpen
  \bibfield  {author} {\bibinfo {author} {\bibfnamefont {T.~A.}\ \bibnamefont
  {Katoh}}, \bibinfo {author} {\bibfnamefont {K.}~\bibnamefont {Ikegami}},
  \bibinfo {author} {\bibfnamefont {N.}~\bibnamefont {Uchida}}, \bibinfo
  {author} {\bibfnamefont {T.}~\bibnamefont {Iwase}}, \bibinfo {author}
  {\bibfnamefont {D.}~\bibnamefont {Nakane}}, \bibinfo {author} {\bibfnamefont
  {T.}~\bibnamefont {Masaike}}, \bibinfo {author} {\bibfnamefont
  {M.}~\bibnamefont {Setou}}, \ and\ \bibinfo {author} {\bibfnamefont
  {T.}~\bibnamefont {Nishizaka}},\ }\href {\doibase 10.1038/s41598-018-33846-5}
  {\bibfield  {journal} {\bibinfo  {journal} {Scientific Reports}\ }\textbf
  {\bibinfo {volume} {8}},\ \bibinfo {pages} {15562} (\bibinfo {year}
  {2018})}\BibitemShut {NoStop}%
\bibitem [{\citenamefont {Brokaw}(1966)}]{Brokaw1966}%
  \BibitemOpen
  \bibfield  {author} {\bibinfo {author} {\bibfnamefont {C.~J.}\ \bibnamefont
  {Brokaw}},\ }\href@noop {} {\bibfield  {journal} {\bibinfo  {journal} {The
  Journal of experimental biology}\ }\textbf {\bibinfo {volume} {45}},\
  \bibinfo {pages} {113} (\bibinfo {year} {1966})}\BibitemShut {NoStop}%
\bibitem [{\citenamefont {Polin}\ \emph {et~al.}(2009)\citenamefont {Polin},
  \citenamefont {Tuval}, \citenamefont {Drescher}, \citenamefont {Gollub},\
  and\ \citenamefont {Goldstein}}]{Polin2009}%
  \BibitemOpen
  \bibfield  {author} {\bibinfo {author} {\bibfnamefont {M.}~\bibnamefont
  {Polin}}, \bibinfo {author} {\bibfnamefont {I.}~\bibnamefont {Tuval}},
  \bibinfo {author} {\bibfnamefont {K.}~\bibnamefont {Drescher}}, \bibinfo
  {author} {\bibfnamefont {J.~P.}\ \bibnamefont {Gollub}}, \ and\ \bibinfo
  {author} {\bibfnamefont {R.~E.}\ \bibnamefont {Goldstein}},\ }\href {\doibase
  10.1126/science.1172667} {\bibfield  {journal} {\bibinfo  {journal}
  {Science}\ }\textbf {\bibinfo {volume} {325}},\ \bibinfo {pages} {487}
  (\bibinfo {year} {2009})}\BibitemShut {NoStop}%
\bibitem [{\citenamefont {Goldstein}\ \emph {et~al.}(2009)\citenamefont
  {Goldstein}, \citenamefont {Polin},\ and\ \citenamefont
  {Tuval}}]{Goldstein2009PRL}%
  \BibitemOpen
  \bibfield  {author} {\bibinfo {author} {\bibfnamefont {R.~E.}\ \bibnamefont
  {Goldstein}}, \bibinfo {author} {\bibfnamefont {M.}~\bibnamefont {Polin}}, \
  and\ \bibinfo {author} {\bibfnamefont {I.}~\bibnamefont {Tuval}},\ }\href
  {\doibase 10.1103/PhysRevLett.103.168103} {\bibfield  {journal} {\bibinfo
  {journal} {Phys. Rev. Lett.}\ }\textbf {\bibinfo {volume} {103}},\ \bibinfo
  {pages} {168103} (\bibinfo {year} {2009})}\BibitemShut {NoStop}%
\bibitem [{\citenamefont {Ma}\ \emph {et~al.}(2014)\citenamefont {Ma},
  \citenamefont {Klindt}, \citenamefont {Riedel-Kruse}, \citenamefont
  {J\"ulicher},\ and\ \citenamefont {Friedrich}}]{Ma2014}%
  \BibitemOpen
  \bibfield  {author} {\bibinfo {author} {\bibfnamefont {R.}~\bibnamefont
  {Ma}}, \bibinfo {author} {\bibfnamefont {G.~S.}\ \bibnamefont {Klindt}},
  \bibinfo {author} {\bibfnamefont {I.~H.}\ \bibnamefont {Riedel-Kruse}},
  \bibinfo {author} {\bibfnamefont {F.}~\bibnamefont {J\"ulicher}}, \ and\
  \bibinfo {author} {\bibfnamefont {B.~M.}\ \bibnamefont {Friedrich}},\ }\href
  {\doibase 10.1103/PhysRevLett.113.048101} {\bibfield  {journal} {\bibinfo
  {journal} {Phys. Rev. Lett.}\ }\textbf {\bibinfo {volume} {113}},\ \bibinfo
  {pages} {048101} (\bibinfo {year} {2014})}\BibitemShut {NoStop}%
\bibitem [{\citenamefont {Wan}\ and\ \citenamefont
  {Goldstein}(2014)}]{Wan2014PRL}%
  \BibitemOpen
  \bibfield  {author} {\bibinfo {author} {\bibfnamefont {K.~Y.}\ \bibnamefont
  {Wan}}\ and\ \bibinfo {author} {\bibfnamefont {R.~E.}\ \bibnamefont
  {Goldstein}},\ }\href {\doibase 10.1103/PhysRevLett.113.238103} {\bibfield
  {journal} {\bibinfo  {journal} {Phys. Rev. Lett.}\ }\textbf {\bibinfo
  {volume} {113}},\ \bibinfo {pages} {238103} (\bibinfo {year}
  {2014})}\BibitemShut {NoStop}%
\bibitem [{\citenamefont {Quaranta}\ \emph {et~al.}(2015)\citenamefont
  {Quaranta}, \citenamefont {Aubin-Tam},\ and\ \citenamefont
  {Tam}}]{Quaranta2015}%
  \BibitemOpen
  \bibfield  {author} {\bibinfo {author} {\bibfnamefont {G.}~\bibnamefont
  {Quaranta}}, \bibinfo {author} {\bibfnamefont {M.-E.}\ \bibnamefont
  {Aubin-Tam}}, \ and\ \bibinfo {author} {\bibfnamefont {D.}~\bibnamefont
  {Tam}},\ }\href@noop {} {\bibfield  {journal} {\bibinfo  {journal} {Physical
  review letters}\ }\textbf {\bibinfo {volume} {115}},\ \bibinfo {pages}
  {238101} (\bibinfo {year} {2015})}\BibitemShut {NoStop}%
\bibitem [{\citenamefont {Ringers}\ \emph {et~al.}(2023)\citenamefont
  {Ringers}, \citenamefont {Bialonski}, \citenamefont {Ege}, \citenamefont
  {Solovev}, \citenamefont {Hansen}, \citenamefont {Jeong}, \citenamefont
  {Friedrich},\ and\ \citenamefont {Jurisch-Yaksi}}]{Ringers2023}%
  \BibitemOpen
  \bibfield  {author} {\bibinfo {author} {\bibfnamefont {C.}~\bibnamefont
  {Ringers}}, \bibinfo {author} {\bibfnamefont {S.}~\bibnamefont {Bialonski}},
  \bibinfo {author} {\bibfnamefont {M.}~\bibnamefont {Ege}}, \bibinfo {author}
  {\bibfnamefont {A.}~\bibnamefont {Solovev}}, \bibinfo {author} {\bibfnamefont
  {J.~N.}\ \bibnamefont {Hansen}}, \bibinfo {author} {\bibfnamefont
  {I.}~\bibnamefont {Jeong}}, \bibinfo {author} {\bibfnamefont {B.~M.}\
  \bibnamefont {Friedrich}}, \ and\ \bibinfo {author} {\bibfnamefont
  {N.}~\bibnamefont {Jurisch-Yaksi}},\ }\href {\doibase 10.7554/eLife.77701}
  {\bibfield  {journal} {\bibinfo  {journal} {eLife}\ }\textbf {\bibinfo
  {volume} {12}},\ \bibinfo {pages} {e77701} (\bibinfo {year}
  {2023})}\BibitemShut {NoStop}%
\end{thebibliography}%

\end{document}